\documentclass[prb,twocolumn,amssymb,amsmath,floatfix,showpacs]{revtex4}
\usepackage{bm,dcolumn,graphicx,hyperref}
\newcommand{\vect}[1]{\mathbf{#1}}
\newcommand{\kp}{\ensuremath{\vect{k} \cdot \vect{p}}}
\newcolumntype{w}[1]{D{.}{.}{#1}}

\begin{document}
\title{Accurate quadratic-response approximation for the self-consistent
       pseudopotential of semiconductor nanostructures}
\author{Bradley A. Foreman}
\email{phbaf@ust.hk}
\affiliation{Department of Physics,
             Hong Kong University of Science and Technology,
             Clear Water Bay, Kowloon, Hong Kong, China}
% \date{\today}

\begin{abstract}
Quadratic-response theory is shown to provide a conceptually simple
but accurate approximation for the self-consistent one-electron
potential of semiconductor nanostructures.  Numerical examples are
presented for GaAs/AlAs and In$_{0.53}$Ga$_{0.47}$As/InP (001)
superlattices using the local-density approximation to
density-functional theory and norm-conserving pseudopotentials without
spin-orbit coupling.  When the reference crystal is chosen to be the
virtual-crystal average of the two bulk constituents, the absolute
error in the quadratic-response potential for $\Gamma_{15}$ valence
electrons is about 2 meV for GaAs/AlAs and 5 meV for
In$_{0.53}$Ga$_{0.47}$As/InP.  Low-order multipole expansions of the
electron density and potential response are shown to be accurate
throughout a small neighborhood of each reciprocal lattice vector,
thus providing a further simplification that is confirmed to be valid
for slowly varying envelope functions.  Although the linear response
is about an order of magnitude larger than the quadratic response, the
quadratic terms are important both quantitatively (if an accuracy of
better than a few tens of meV is desired) and qualitatively (due to
their different symmetry and long-range dipole effects).
\end{abstract}


\maketitle

\section{Introduction}

The potential energy of the electron in the single-electron
approximation provides the theoretical foundation for most modeling
and interpretation of the electronic energy bands of nonmagnetic
semiconductors.  Implementations of this concept range in
sophistication from simple empirical pseudopotentials
\cite{CoBe66,ChCo76} to first-principles quasiparticle
self-energies. \cite{Aulbur00,Onida02} Modern research on
semiconductor physics deals mainly with complex systems (such as
nanostructures and novel materials), although the spintronic
properties of traditional semiconductors are also of strong current
interest.  It is therefore somewhat surprising that, in this day of
supposedly advanced understanding, no clear physical picture exists of
something so basic as the one-electron potential energy at an ideal
GaAs/AlAs or GaInAs/InP heterojunction.

The key issue is the self-consistent redistribution of charge that
must occur at any interface between two bulk semiconductors.
First-principles numerical calculations based on density-functional
theory and norm-conserving pseudopotentials \cite{Payne92,Mart04} are
now routinely performed on small systems (say, a few tens of atoms)
using personal computers and free
software. \cite{Gonze02,Gonze05,ABINIT} However, such computations are
still not feasible for large quantum dots and quantum wires.  Even if
they were, there is no means at present for separating this
information from a huge numerical calculation.  That is, no one has
yet demonstrated an accurate method for breaking down the
self-consistent potential into simple conceptual units that can be
used to {\em understand} the interface potential and model it using
less numerically intensive methods (such as empirical pseudopotentials
or envelope-function theory).

The purpose of this paper is to demonstrate that the
quadratic-response formalism developed in Refs.\ \onlinecite{Fore05a}
and \onlinecite{Fore05b} provides a conceptually simple method for
constructing the interface potential that is accurate to within a few
meV in typical semiconductor heterostructures.  In this approach, the
heterostructure is treated as a perturbation of a bulk reference
crystal, and the self-consistent heterostructure potential is built up
as a series of successively smaller corrections to the bulk potential.

The leading term is the linear response, which has the form of a
superposition of elementary quasi-atomic building blocks.  This form
is widely used in empirical pseudopotential models (see the literature
review in Sec.\ \ref{sec:literature}), but it is shown here to be
accurate to only a few tens of meV.  The quadratic response is then
added in the form of {\em diatomic} building blocks, which are
qualitatively different because they carry a dipole moment even in
cubic semiconductors.  Numerical implementations of the theory are
presented here for the paradigmatic common-atom and no-common-atom
systems GaAs/AlAs and In$_{0.53}$Ga$_{0.47}$As/InP.  The
quadratic-response approximation for $\Gamma_{15}$ valence electrons
is shown to be accurate to within about 2 meV for GaAs/AlAs and 5 meV
for In$_{0.53}$Ga$_{0.47}$As/InP.

In typical heterostructures, the low-energy electron and hole states
can be represented by slowly varying envelope functions.  The
properties of such states are determined by the pseudopotential not
just at the bulk reciprocal-lattice vectors $\vect{G}$ (as would be
the case in a bulk semiconductor), but throughout a small
$\vect{k}$-space neighborhood of each $\vect{G}$.  Within such a
neighborhood, an analytic function of $\vect{k}$ can be approximated
using a power series (which is equivalent to introducing a multipole
expansion of a localized function in coordinate space).  This paper
also examines the accuracy of such power series and demonstrates that
the truncated expansions introduced in Refs.\ \onlinecite{Fore05a} and
\onlinecite{Fore05b} are sufficiently accurate for slowly varying
envelope functions.  The application of these multipole expansions in
actual envelope-function calculations is presented in another
paper. \cite{Fore07b}

The paper begins in Sec.\ \ref{sec:literature} with a review of
previous models of the interface potential.  Section \ref{sec:abinit}
describes the choices made in defining the model system used here for
numerical calculations.  The basic features of the self-consistent
density and potential in GaAs/AlAs and In$_{0.53}$Ga$_{0.47}$As/InP
superlattices are examined in Sec.\ \ref{sec:VBO}; here the
valence-band offsets calculated for the model systems are found to be
in reasonably good agreement with experiment.  Section
\ref{sec:response} outlines the methods used to calculate the linear
and quadratic response to the heterostructure perturbation and
discusses the physical significance of the results.  Localization of
the density and potential response and the use of multipole expansions
at small wave vectors are considered in Sec.\ \ref{sec:multipole}.
Section \ref{sec:quaderr} shows that the cumulative errors arising
from the quadratic-response and truncated-multipole approximations are
limited to a few meV for the systems studied.  The conclusions of the
paper are reviewed in Sec.\ \ref{sec:conclusions}.  Finally, Sec.\
\ref{sec:limitations} discusses several limitations of the
calculations presented here, along with possible extensions of this
work that could be used to overcome some of these limitations.

\section{Previous descriptions of the interface potential}

\label{sec:literature}

Two simple non-self-consistent models of the heterostructure
pseudopotential are widely used.  The simplest is the {\em planar
interface} model, in which the potential is assumed to make an abrupt
transition from one periodic bulk potential to another at some
predefined interface plane.
\cite{MarInk84,MaSm86,BrHu87,Xia89,SmMa90,CuVH92,DenTil99,DenTil02}
This requires as input only the bulk pseudopotential form factors.
\cite{CoBe66,ChCo76} The assumption of a planar interface provides, in
essence, a simple recipe for interpolating the pseudopotential in
$\vect{k}$ space between the bulk reciprocal-lattice vectors
$\vect{G}$.  A discontinuous potential is, however, clearly far from
self-consistent (since by Poisson's equation it would require an
infinite electron density), and a planar interface fails to account
for the basic physics of the bonds between atoms at a no-common-atom
interface. \cite{MagZun03}

An arguably more realistic potential (although not all would
agree\cite{DenTil02}) is provided by the {\em atomic superposition}
model, which is defined as a linear combination of screened atomic
pseudo\-po\-ten\-tials.
\cite{AndBalCar78,AndCar80,Xia88a,Fried89,MaZu94,WaZu95,FuZu97} Here
the atomic potentials are assumed to be known as continuous functions
of $\vect{k}$; \cite{AniHei65} hence, no interpolation is required.
In its simplest form this model may not accurately reproduce
differences between bulk semiconductors, \cite{MaZu94} since (for
example) the screened As pseudopotential obtained by fitting the
properties of bulk GaAs will generally differ from that obtained by
fitting AlAs.

However, M\"ader and Zunger have overcome this difficulty by using
environment-dependent pseudo\-po\-ten\-tials, \cite{MaZu94} in which
the replacement of one Ga atom in GaAs by Al is accompanied by small
changes in each of the four neighboring As pseudopotentials.  The
atomic pseudopotential and its associated nearest-neighbor
perturbations may be grouped conceptually into a single quasi-atomic
unit, which has the site symmetry $T_d$ of an Al impurity in GaAs
rather than the spherical symmetry of an isolated Al atom.  Thus, in
this more general {\em quasi-atomic superposition} model,
\cite{MaZu94} the heterostructure potential is constructed as a linear
combination of quasi-atomic potentials, each of which has the site
symmetry of a substitutional impurity.  (The planar interface model
may also be interpreted as a superposition of elementary units,
\cite{Klein81} but in this case the units do not have the atomic site
symmetry.)

The superposition potential is also manifestly non-self-consistent, if
only because the exchange-correlation potential is nonlinear.  The
assumption of linearity has important physical consequences.  As shown
in Eq.\ (10) of Ref.\ \onlinecite{WaZu97}, one of these is the absence
of direct $\Gamma_1$--$X_1$ intervalley coupling in GaAs/AlAs (001)
heterostructures.  The same conclusion was reached in Ref.\
\onlinecite{Fore98b} using a superposition of quasi-atomic potentials
that were constructed by a different method.  These conclusions were
criticized by Takhtamirov and Volkov \cite{TakVol00a} on the grounds
that the true interface potential does not have the assumed form, and
that there is consequently no basis for assuming $\Gamma_1$--$X_1$
coupling to be any smaller than $\Gamma_1$--$X_3$ coupling (the latter
of which is nonvanishing in the quasi-atomic superposition
model\cite{WaZu97,Fore98b}).  This criticism is, of course, entirely
correct if the self-consistent interface potential is viewed as
completely unknown.

Indeed, many derivations of envelope-function models are prefaced by a
warning that the potential within a few lattice constants of the
interface is too complicated for direct analysis,
\cite{Bast91,Laik92,KiGeLu98,Vasko99,Rod02,Rod06} even in the case of
an ideal interface (i.e., no interface roughness, defects, or
interdiffusion).  In early works it was sometimes concluded that the
details of the interface potential have no significant influence on
slowly varying envelope functions.  Today, however, it is widely
accepted that these details are of critical importance in determining
the strength of small symmetry-breaking interface terms in the
Hamiltonian, such as the above-mentioned intervalley mixing,
\cite{WaZu97,Fore98b,TakVol00a} valence-band mixing,
\cite{Fore98b,IvKaRo96,TakVol99b} and conduction- and valence-band
Rashba coupling \cite{Wink03} (see the following paper \cite{Fore07b}
for further study of some of these topics).  But the assumption that
the self-consistent potential cannot be understood in simple terms
remains unchallenged by most envelope-function theorists.

This problem was tackled in a landmark series of papers by Baroni {\em
et al}., who showed that linear-response theory is very successful in
predicting valence-band offsets in semiconductor
het\-ero\-struc\-tures.
\cite{ResBarBal89,BaReBa88,BaReBaPe89,PBBR90,PBRB91,%
ColResBar91,BaPeReBa92,PerBar94,PeMoBaMo96,MonPer96} In this approach,
the ionic pseudopotential of the heterostructure is treated as a
perturbation of a bulk reference crystal.  The linear response to this
perturbation was shown to give accurate predictions for the
valence-band offset (within about 0.01 eV of the value obtained from a
direct calculation, and in good agreement with experiment) if the
reference crystal was chosen to be the virtual-crystal average of the
two bulk materials, since the quadratic response is then the same on
both sides of the heterojunction and gives no contribution to the
offset.  However, if one is interested in the actual
position-dependent potential, including the details of its behavior
near the interface, the quadratic response is not generally
negligible.

Nonlinear screened empirical pseudopotentials were introduced by Magri
and Zunger, \cite{MagZun02} who extended the environment-dependent
concept of M\"ader and Zunger \cite{MaZu94} to the no-common-atom
InAs/GaSb system.  Nonlinearity arises in this case because the
nearest-neighbor perturbations depend on which atoms occupy the
neighboring sites.  The pseudopotentials can therefore no longer be
interpreted as purely quasi-atomic units; a complete description
requires the inclusion of nearest-neighbor {\em diatomic} building
blocks as well. \cite{note:MagZun02} This represents an important
conceptual advance in the modeling of heterostructure
pseudopotentials.

The present work develops this concept more thoroughly by extending
the first-principles linear-response theory of Baroni {\em et
al}.\cite{BaReBaPe89} to include the diatomic quadratic response for
arbitrary pairs of atoms.  The results show that the environment
dependence of the empirical pseudopotentials in Refs.\
\onlinecite{MaZu94} and \onlinecite{MagZun02} is qualitatively
incomplete, since (for example) the leading contribution to
$\Gamma_1$--$X_1$ mixing in GaAs/AlAs (001) superlattices arises from
second-nearest-neighbor Ga--Al pairs.  (However, this contribution is
about two orders of magnitude smaller than the linear response, which
supports the conclusion that $\Gamma_1$--$X_1$ mixing is much weaker
than $\Gamma_1$--$X_3$ mixing.)  Also, the diatomic nearest-neighbor
response for no-common-atom systems calculated here has a dipole
moment that generates long-range electric fields extending over the
entire superlattice.  Such fields do not occur in the empirical
pseudopotential interpolation scheme of Magri and Zunger,
\cite{MagZun02} which produces only short-range potentials.  This
suggests that a more complete description of quadratic diatomic
response may be of benefit in achieving more realistic empirical
pseudopotentials.

\section{Numerical considerations}

\label{sec:abinit}

\subsection{Definition of model system}

\label{subsec:choices}

All of the numerical results in this paper are derived from plane-wave
pseudopotential total-energy calculations \cite{Payne92,Mart04}
performed using the \textsc{abinit}
software. \cite{Gonze02,Gonze05,ABINIT} This package provides a
variety of options, but the particular calculations reported here used
the local density approximation (LDA) to density-functional theory and
the nonlocal norm-conserving pseudopotentials of Hartwigsen,
Goedecker, and Hutter \cite{HaGoHu98} with no spin-orbit coupling.

Choosing a particular physical structure and a particular set of
technical ingredients (such as plane-wave kinetic-energy cutoffs and
$\vect{k}$-point sampling) defines a model system whose properties can
be calculated self-consistently to an accuracy approaching that of
machine precision.  These ``exact'' model calculations are used here
and in the following paper \cite{Fore07b} as a benchmark for
comparison with various approximations used in the construction of the
first-principles envelope-function theory of Ref.\
\onlinecite{Fore05b}.  The main objective of this paper is to pin down
as closely as possible how much error arises from the
quadratic-response approximation and the multipole expansions of the
linear and quadratic response.  The approximate potentials derived
here are also used directly in a subsequent paper \cite{Fore07b} that
examines the accuracy of first-principles envelope-function theory in
superlattices.

With these goals in mind, relatively low kinetic-energy cutoffs (see
below) and numbers of special $\vect{k}$ points were chosen in order
to make the ``exact'' calculations feasible for the large
superlattices in which envelope-function theory is valid.  In
particular, two independent $\vect{k}$ points were used for GaAs/AlAs
superlattices with $D_{2d}$ symmetry, while four points were used for
In$_{0.53}$Ga$_{0.47}$As/InP superlattices with $C_{2v}$ symmetry.
The total energy is not completely converged at these values, but the
valence-band offsets calculated here are nevertheless in good
agreement with experiment and with previous first-principles
calculations reported in the literature.

\begin{figure}
  \includegraphics[width=8.5cm,clip]{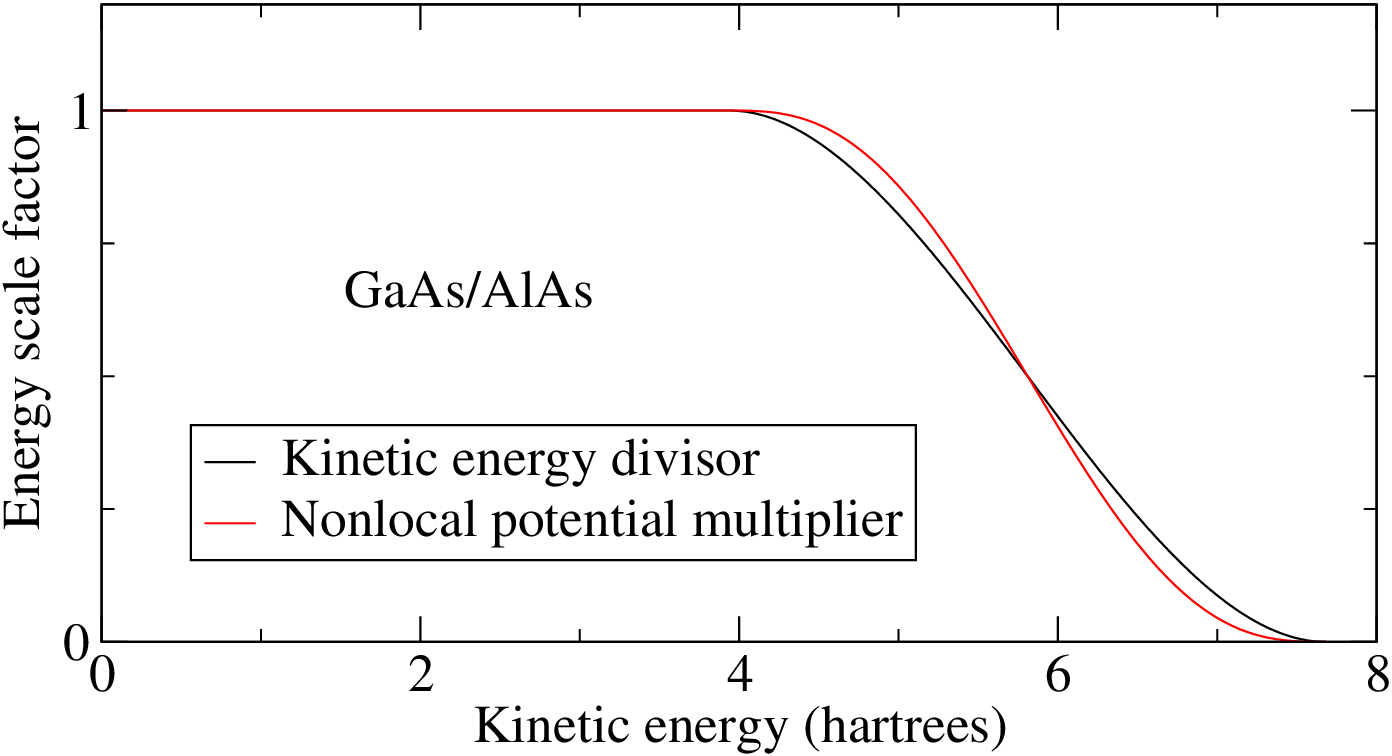}
  \caption{\label{fig:GaAs_kinetic_divisor} (Color online) Kinetic
  energy divisor (with continuous slope) and nonlocal potential energy
  multiplier (with continuous curvature) used to smooth out
  discontinuities in the momentum-space electron density and
  pseudopotential.  The multiplier used for the local pseudopotential
  has the same shape, but the energy scale is four times that shown.}
\end{figure}

In order to calculate multipole moments of the linear and quadratic
response, it is necessary that the electron density and short-range
potential response be well localized within the given large
supercells.  However, sharp cutoffs in the plane-wave basis produce
spurious long-range Gibbs oscillations that are very small but
nonetheless large enough to completely swamp the physical multipole
values.  To suppress these oscillations, the smoothing functions shown
in Fig.\ \ref{fig:GaAs_kinetic_divisor}
were applied to the kinetic and potential energy.  The chosen
kinetic-energy cutoff (i.e., the value at which the divisor in Fig.\
\ref{fig:GaAs_kinetic_divisor} goes to zero) includes 283 plane waves
at the $\Gamma$ point in bulk material (113 of which are not altered
by smoothing) and more than 14000 plane waves in a 96-atom
superlattice.

\subsection{Conventions defining the absolute energy}

\label{subsec:conventions}

There are two conventions in the momentum-space total-energy formalism
that determine the absolute value of the local potential. First, the
mean value of the Hartree potential is set to zero by definition:
\cite{Ihm79,YinCoh82a,YinCoh82b}
\begin{equation}
  V_{\mathrm{Har}}(\vect{q}) = (1 - \delta_{\vect{q} \vect{0}})
  \frac{4 \pi}{q^2} n(\vect{q}) \label{eq:V_Har}
\end{equation}
(where $n$ is the valence electron density), because in a neutral
system the net contribution to the total energy from the Coulomb
interaction terms (for the electron-ion system) at wave vector
$\vect{q} = \vect{0}$ is identically zero.

A similar convention \cite{Ihm79} is often used for the mean value of
the short-range part of the local pseudopotential.  This is given by
$\bar{V}_{\text{psp}} = \sum_j \bar{V}_j$, where
\cite{Ihm79,YinCoh82a,Payne92,Mart04}
\begin{equation}
  \bar{V}_j = \frac{1}{\Omega} \int \left( V_j (r) + \frac{Z_j}{r}
  \right) d^3 r ,
\end{equation}
in which $\Omega$ is the unit cell volume and $V_j$ and $Z_j$ are the
local pseudopotential and valence charge of ion $j$.  The contribution
from $\bar{V}_{\text{psp}}$ to the total energy is just the constant
$E_{\text{core}} = Z \Omega \bar{V}_{\text{psp}}$, where $Z = \sum_j
Z_j$.  Hence, it is permissible to set $\bar{V}_{\text{psp}}$ to zero
in the Kohn-Sham equations. \cite{Ihm79}

However, in order to calculate the nonlinear response correctly,
$\bar{V}_{\text{psp}}$ must be included in the local pseudopotential.
In the present work, $\bar{V}_{\text{psp}}$ was added to both the
local pseudopotential and the energy eigenvalues after the conclusion
of each self-consistent calculation.  Therefore, the absolute energy
scale for all figures in this paper (and the next\cite{Fore07b}) is
defined solely by the convention of setting the mean Hartree potential
to zero.

\section{Valence-band offsets in superlattices}

\label{sec:VBO}

\begin{figure}
  \includegraphics[width=8.5cm,clip]{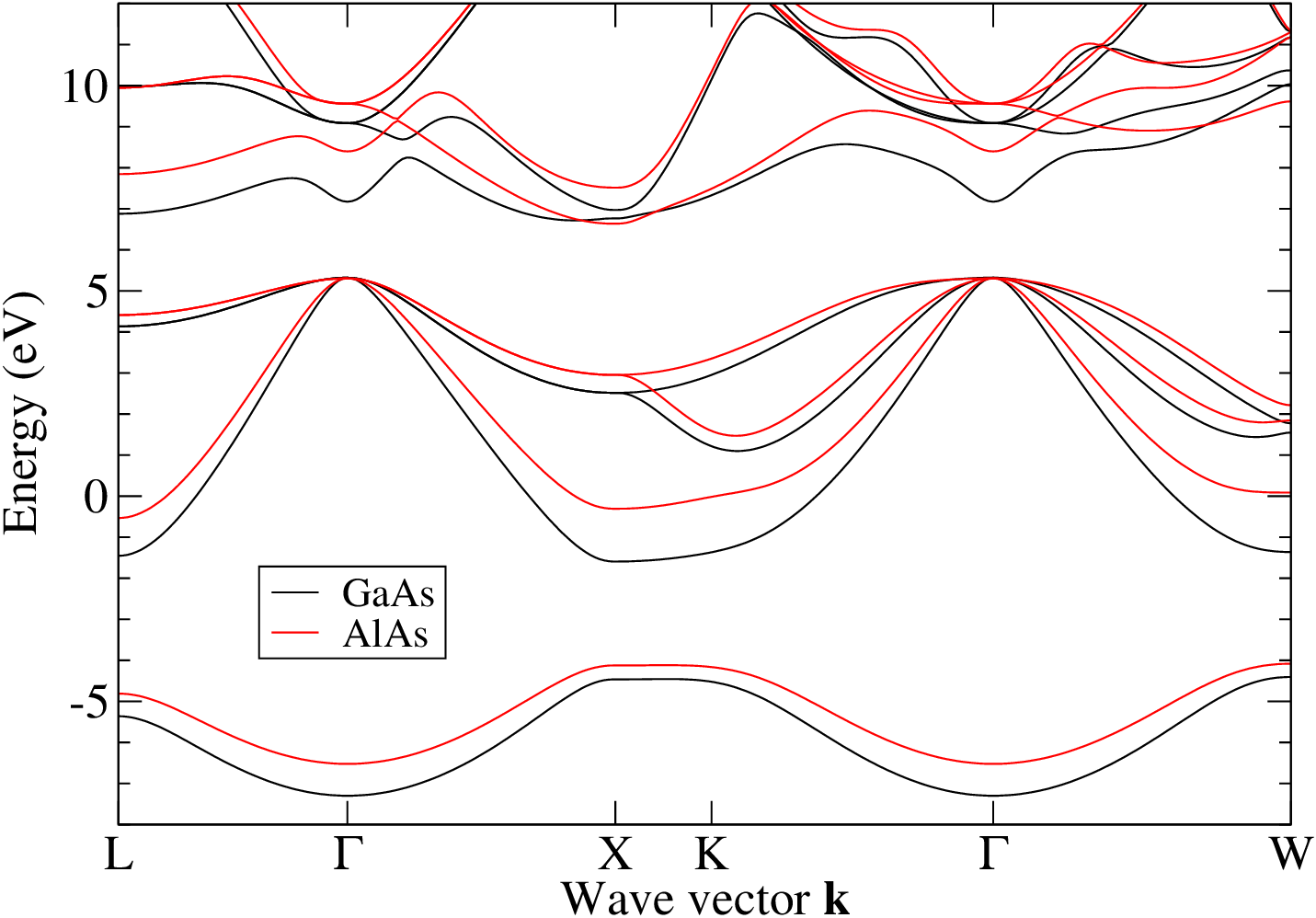}
  \caption{\label{fig:GaAs_bulk_bs_AB} (Color online) Energy band
  structure of bulk GaAs and AlAs.}
\end{figure}

\begin{figure}
  \includegraphics[width=8.5cm,clip]{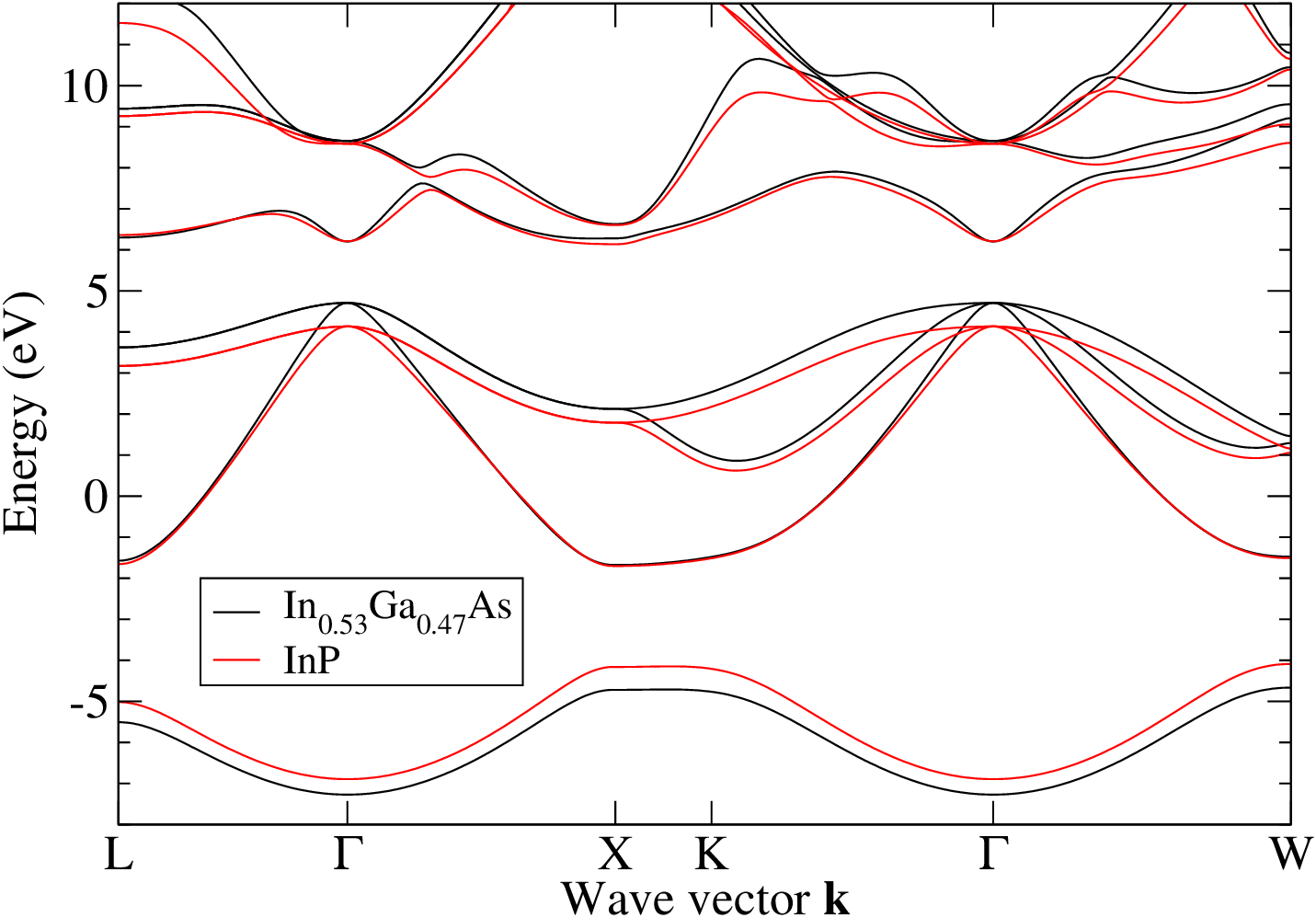}
  \caption{\label{fig:InP_bulk_bs_AB} (Color online) Energy band
  structure of bulk In$_{0.53}$Ga$_{0.47}$As and InP.}
\end{figure}

Energy band structures for the model system described in Sec.\
\ref{sec:abinit} are presented in Figs.\ \ref{fig:GaAs_bulk_bs_AB}
and \ref{fig:InP_bulk_bs_AB} for the bulk materials GaAs, AlAs,
In$_{0.53}$Ga$_{0.47}$As, and InP\@.
The conduction bands in these figures are clearly not realistic, as
the minimum at $L$ lies slightly below the $\Gamma$ minimum in GaAs,
and is only slightly above it in In$_{0.53}$Ga$_{0.47}$As and InP\@.
The calculated $\Gamma_{1\mathrm{c}}$--$\Gamma_{15\mathrm{v}}$ energy
gaps are 1.850 eV for GaAs and 1.492 eV for In$_{0.53}$Ga$_{0.47}$As,
which are 13\% and 61\% larger than the experimental values
\cite{VurMeyRM01} (with spin-orbit splitting removed) of 1.633 eV and
0.925 eV\@.  Nevertheless, the model does capture many of the gross
qualitative trends in experimentally determined band structures,
\cite{ChCo76} such as the predominance of the $X$ valley in AlAs
relative to GaAs.

\begin{figure}
  \includegraphics[width=8.5cm,clip]{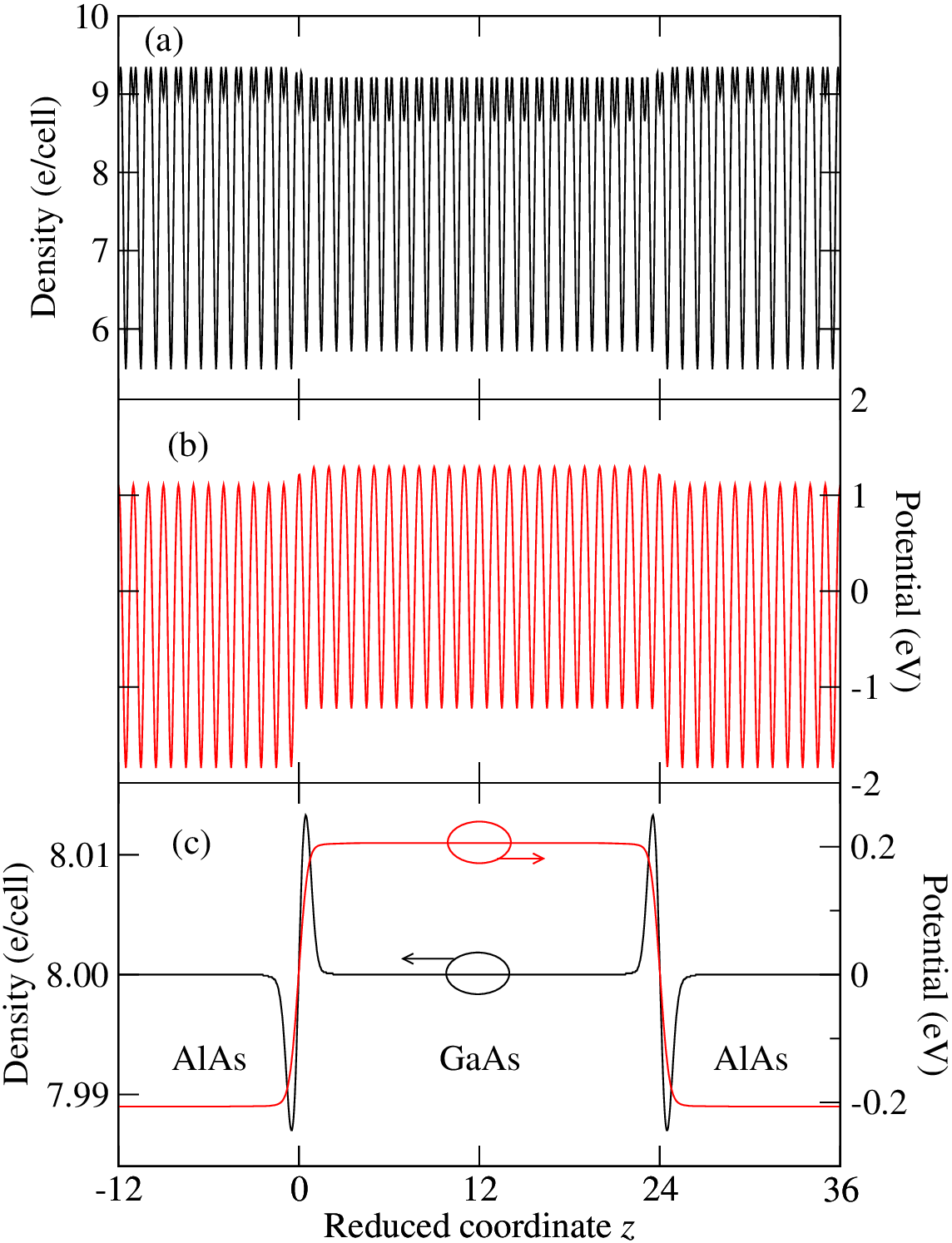}
  \caption{\label{fig:GaAs_slat_overview} (Color online) Density and
  potential of a (001) (GaAs)$_{24}$(AlAs)$_{24}$ superlattice: (a)
  Planar average of electron density. (b) Planar average of Hartree
  potential. (c) Macroscopic average of density and potential.}
\end{figure}

\begin{figure}
  \includegraphics[width=8.5cm,clip]{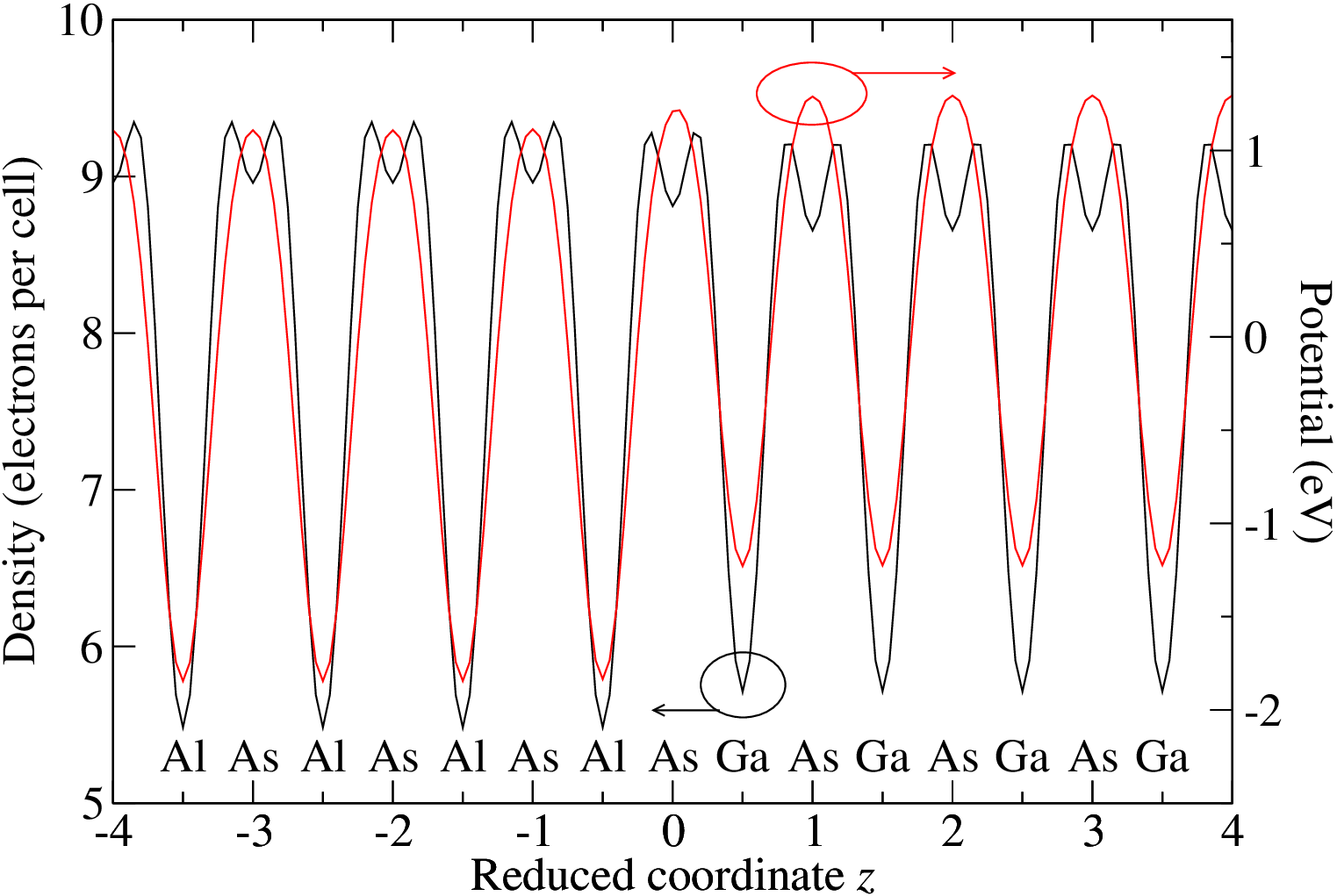}
  \caption{\label{fig:GaAs_slat_closeup} (Color online) Expanded view
  of the planar average density and potential from Fig.\
  \ref{fig:GaAs_slat_overview}.}
\end{figure}

\begin{figure}
  \includegraphics[width=8.5cm,clip]{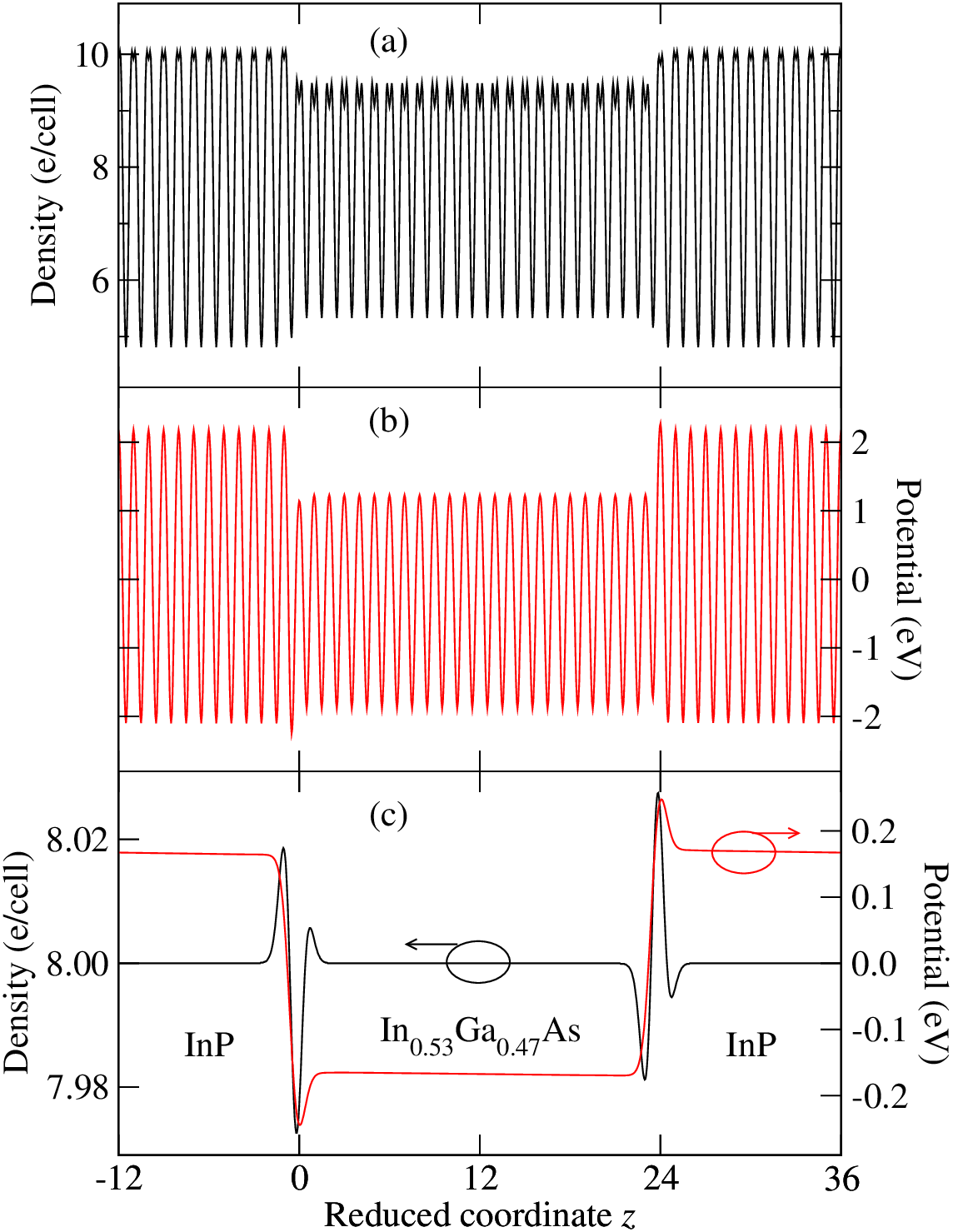}
  \caption{\label{fig:InP_slat_overview} (Color online) Density and
  potential of a (001) (In$_{0.53}$Ga$_{0.47}$As)$_{24}$(InP)$_{24}$
  superlattice: (a) Planar average of electron density. (b) Planar
  average of Hartree potential. (c) Macroscopic average of density and
  potential.}
\end{figure}

\begin{figure}
  \includegraphics[width=8.5cm,clip]{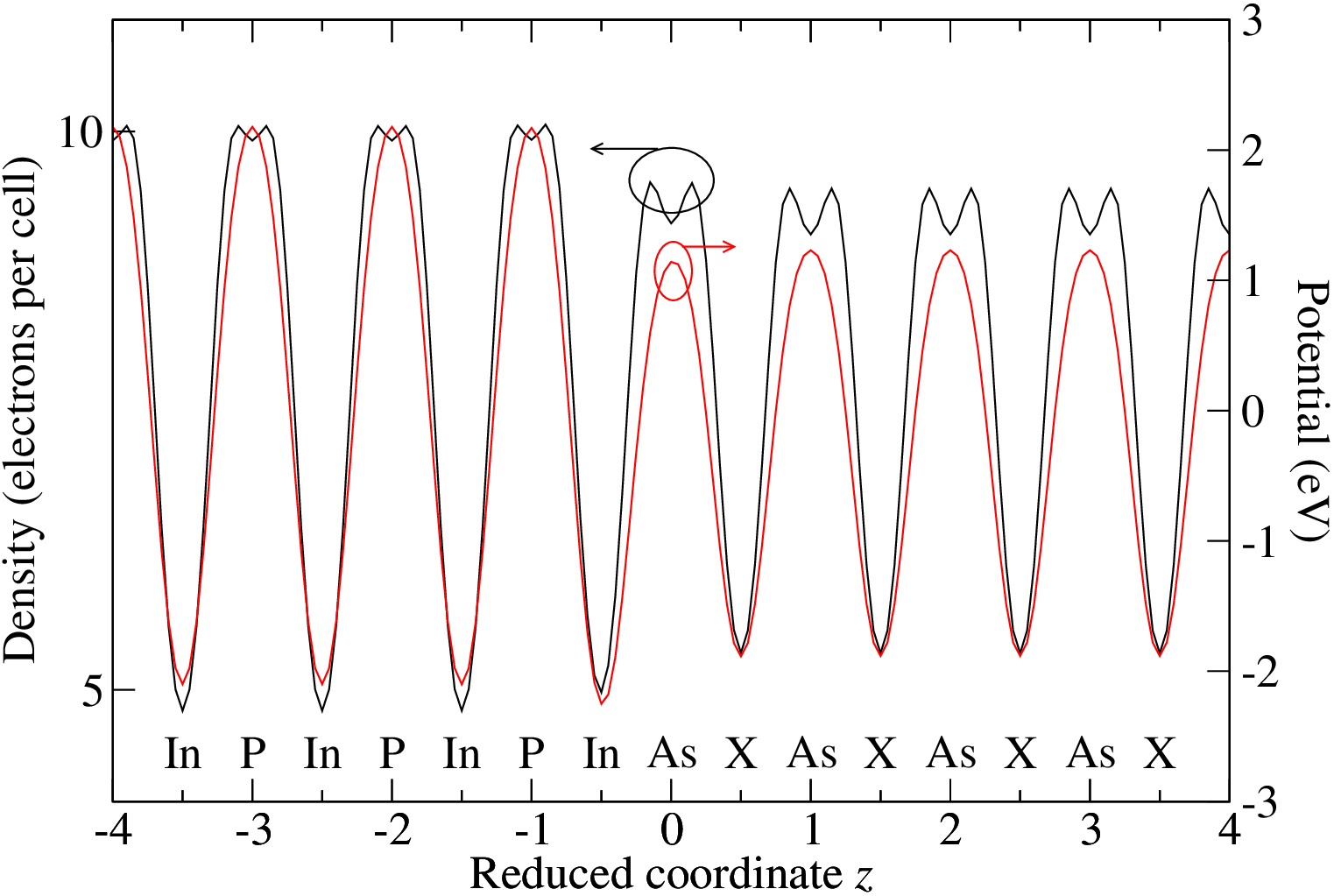}
  \caption{\label{fig:InP_slat_closeup} (Color online) Expanded view
  of the planar average density and potential from Fig.\
  \ref{fig:InP_slat_overview}.  Here X represents the virtual atom
  In$_{0.53}$Ga$_{0.47}$.}
\end{figure}

It is important to note that the difference between the valence-band
maxima in Figs.\ \ref{fig:GaAs_bulk_bs_AB} and
\ref{fig:InP_bulk_bs_AB} is not equal to the valence-band offset.
\cite{BaReBaPe89} To obtain the valence-band offset, one must add to
this difference the shift in the macroscopic Hartree potential of the
two materials as given by a supercell calculation. \cite{BaReBaPe89}
The latter shift is shown in Figs.\ \ref{fig:GaAs_slat_overview} and
\ref{fig:GaAs_slat_closeup} for a GaAs/AlAs superlattice%
and in Figs.\ \ref{fig:InP_slat_overview} and
\ref{fig:InP_slat_closeup} for an In$_{0.53}$Ga$_{0.47}$As/InP
superlattice.
These figures show planar averages and macroscopic averages
\cite{BaBaRe88,Jackson99,Rob73} of the electron density and Hartree
potential, with distance given in units of the monolayer spacing $d =
\frac12 a$, where $a$ is the cubic lattice constant.

To calculate the macroscopic average, the microscopic functions were
first averaged over the volume of a bulk unit cell, as in Ref.\
\onlinecite{BaBaRe88}.  Then, to ensure that the output is smooth and
finite even when the input is singular (e.g., in the case of a
truncated multipole expansion), an additional averaging was performed
with respect to the normalized Gaussian function \cite{Jackson99}
\begin{subequations} \label{eq:macroave}
\begin{equation}
  f(z) = \frac{1}{\sigma} \exp [-\pi (z / \sigma)^2] ,
  \label{eq:macroave_z}
\end{equation}
which is equivalent to multiplying in momentum space by
\begin{equation}
  g(k) = \exp [-\pi (k \sigma / 2\pi)^2] . \label{eq:macroave_k}
\end{equation}
\end{subequations}
In Figs.\ \ref{fig:GaAs_slat_overview}(c) and
\ref{fig:InP_slat_overview}(c) the distance $\sigma$ was chosen to be
$\sigma = d$.

The macroscopic density for GaAs/AlAs in Fig.\
\ref{fig:GaAs_slat_overview}(c) has the familiar interface dipole
shape, \cite{BaBaRe88,BaReBaPe89} leading to a steplike shift in the
Hartree potential of 0.412 eV\@.  The shift predicted by linear
response theory is very close to this value, at 0.416 eV (see below
for a more detailed study of the error in this approximation).  These
results agree well with the shift of 0.41 eV calculated in Ref.\
\onlinecite{BaReBaPe89}.  Since the difference between the
valence-band maxima in Fig.\ \ref{fig:GaAs_bulk_bs_AB} is 18 meV, the
valence-band offset for a (001) GaAs/AlAs heterojunction is 0.430 eV
for the model system, as compared to the value of 0.45 eV reported in
Ref.\ \onlinecite{BaReBaPe89}.  If this figure is corrected for
spin-orbit coupling and quasiparticle effects using the values in
Ref.\ \onlinecite{BaReBaPe89}, the net offset is roughly 0.56 eV, in
good agreement with the experimental value \cite{VurMeyRM01} of 0.53
eV\@.

The macroscopic density of the no-common-atom
In$_{0.53}$Ga$_{0.47}$As/InP superlattice in Fig.\
\ref{fig:InP_slat_overview}(c) does not have a simple dipole shape
because (to leading order) it is a superposition of two offset
dipoles, one for the cation and one for the
anion. \cite{BaReBaPe89,PBBR90} As discussed below, there is also a
slight asymmetry of the two interfaces, leading to a macroscopic
electric field that is barely visible in Fig.\
\ref{fig:InP_slat_overview}(c).  The linear Hartree shifts [from Eq.\
(\ref{eq:addHart0}) with $l = 0$] for cations and anions are $+0.366$
eV and $-0.697$ eV, as compared to the values of $+0.34$ eV and
$-0.58$ eV reported in Refs.\ \onlinecite{BaReBaPe89} and
\onlinecite{PBBR90}.  The net linear Hartree shift of $-0.331$ eV
agrees well with the mean shift of $-0.334$ eV in Fig.\
\ref{fig:InP_slat_overview}(c).  Combining this with the 0.574 eV
difference in the valence-band maxima of Fig.\
\ref{fig:InP_bulk_bs_AB}, the net valence-band offset for the model
In$_{0.53}$Ga$_{0.47}$As/InP system is 0.240 eV\@.  After adjusting
for experimental spin-orbit splitting, \cite{VurMeyRM01} the
calculated offset of 0.313 eV is close to the experimental value
\cite{VurMeyRM01} of 0.345 eV\@.

\section{Linear and quadratic response}

\label{sec:response}

Having established that the model system adopted here provides at
least a rough approximation of physical reality, the next step is to
calculate the linear and quadratic response to virtual-crystal
perturbations of a bulk reference crystal.  The reference crystal is
chosen here to be the virtual-crystal average of the bulk constituents
(i.e., Al$_{0.5}$Ga$_{0.5}$As for GaAs/AlAs and
In$_{0.765}$Ga$_{0.235}$As$_{0.5}$P$_{0.5}$ for
In$_{0.53}$Ga$_{0.47}$As/InP).  The lattice constant for all
calculations was fixed at the value obtained by minimizing the total
energy of the reference crystal (i.e., $a = 10.5$ bohr for
Al$_{0.5}$Ga$_{0.5}$As and $a = 10.9$ bohr for
In$_{0.765}$Ga$_{0.235}$As$_{0.5}$P$_{0.5}$).

The perturbation of the heterostructure relative to the reference
crystal is defined by the change in pseudopotential
\begin{equation}
  \Delta V_{\mathrm{psp}} (\vect{x}) = \sum_{\alpha} \sum_{\vect{R}}
  \theta^{\alpha}_{\vect{R}} v^{\alpha}_{\mathrm{ion}} (\vect{x} -
  \vect{R}_{\alpha}) ,
\end{equation}
which is written as a local potential for simplicity (see Ref.\
\onlinecite{Fore05b} for the nonlocal case).  Here
$v^{\alpha}_{\mathrm{ion}}(\vect{x})$ is the ionic pseudopotential of
atom $\alpha$, $\vect{R}_{\alpha}$ is the position of atom $\alpha$ in
unit cell $\vect{R}$, and $\theta^{\alpha}_{\vect{R}}$ is the change
in fractional weight of atom $\alpha$ in cell $\vect{R}$ of the
heterostructure relative to the reference crystal.  Since the total
change in fractional weight at each site must add to zero, this can be
rewritten as \cite{Fore05b}
\begin{equation}
  \Delta V_{\mathrm{psp}} (\vect{x}) = \sideset{}{'} \sum_{\alpha}
  \sum_{\vect{R}} \theta^{\alpha}_{\vect{R}} \Delta
  v^{\alpha}_{\mathrm{ion}} (\vect{x} - \vect{R}_{\alpha}) ,
\end{equation}
where the sum covers only independent values of $\alpha$ (e.g., either
Ga or Al---but not both---in GaAs/AlAs), and $\Delta
v^{\alpha}_{\mathrm{ion}}$ is the change in
$v^{\alpha}_{\mathrm{ion}}$ relative to the reference crystal.

If $n(\vect{x})$ is the exact density of the heterostructure, the
linear and quadratic response to virtual perturbations
$\theta^{\alpha}_{\vect{R}}$ are defined as the derivatives
\begin{subequations} \label{eq:n1n2}
\begin{align}
  \Delta n^{\alpha}_{\vect{R}} (\vect{x}) & = \frac{\partial
  n(\vect{x})}{\partial \theta^{\alpha}_{\vect{R}}} , \\ \Delta
  n^{\alpha\alpha'}_{\vect{R}\vect{R}'} (\vect{x}) & = \frac12
  \frac{\partial^2 n(\vect{x})}{\partial \theta^{\alpha}_{\vect{R}}
  \partial \theta^{\alpha'}_{\vect{R}'}} .
\end{align}
\end{subequations}
The total density may then be reconstructed from the power series
\begin{subequations} \label{eq:nonlinear}
\begin{align}
  n(\vect{x}) & = n^{(0)}(\vect{x}) + n^{(1)}(\vect{x}) +
  n^{(2)}(\vect{x}) + \cdots , \\ n^{(1)}(\vect{x}) & = \sideset{}{'}
  \sum_{\alpha,\vect{R}} \theta^{\alpha}_{\vect{R}} \Delta
  n^{\alpha}_{\vect{R}} (\vect{x}) , \\ n^{(2)}(\vect{x}) & =
  \sideset{}{'} \sum_{\alpha,\vect{R}} \sideset{}{'}
  \sum_{\alpha',\vect{R}'} \theta^{\alpha}_{\vect{R}}
  \theta^{\alpha'}_{\vect{R}'} \Delta
  n^{\alpha\alpha'}_{\vect{R}\vect{R}'} (\vect{x}) , \label{eq:n2}
\end{align}
\end{subequations}
where $n^{(0)}(\vect{x})$ is the density of the reference crystal.  In
the present work, this power series was truncated at the second order.
The derivatives (\ref{eq:n1n2}) were calculated by the direct
supercell method, \cite{ResBarBal89} in which (for example) the
Al$_{0.5}$Ga$_{0.5}$As reference crystal was perturbed by replacing
one or two Al$_{0.5}$Ga$_{0.5}$ atoms with either
Al$_{0.55}$Ga$_{0.45}$ or Al$_{0.45}$Ga$_{0.55}$.  See the Appendix
for further details.

\begin{figure}
  \includegraphics[width=8.5cm,clip]{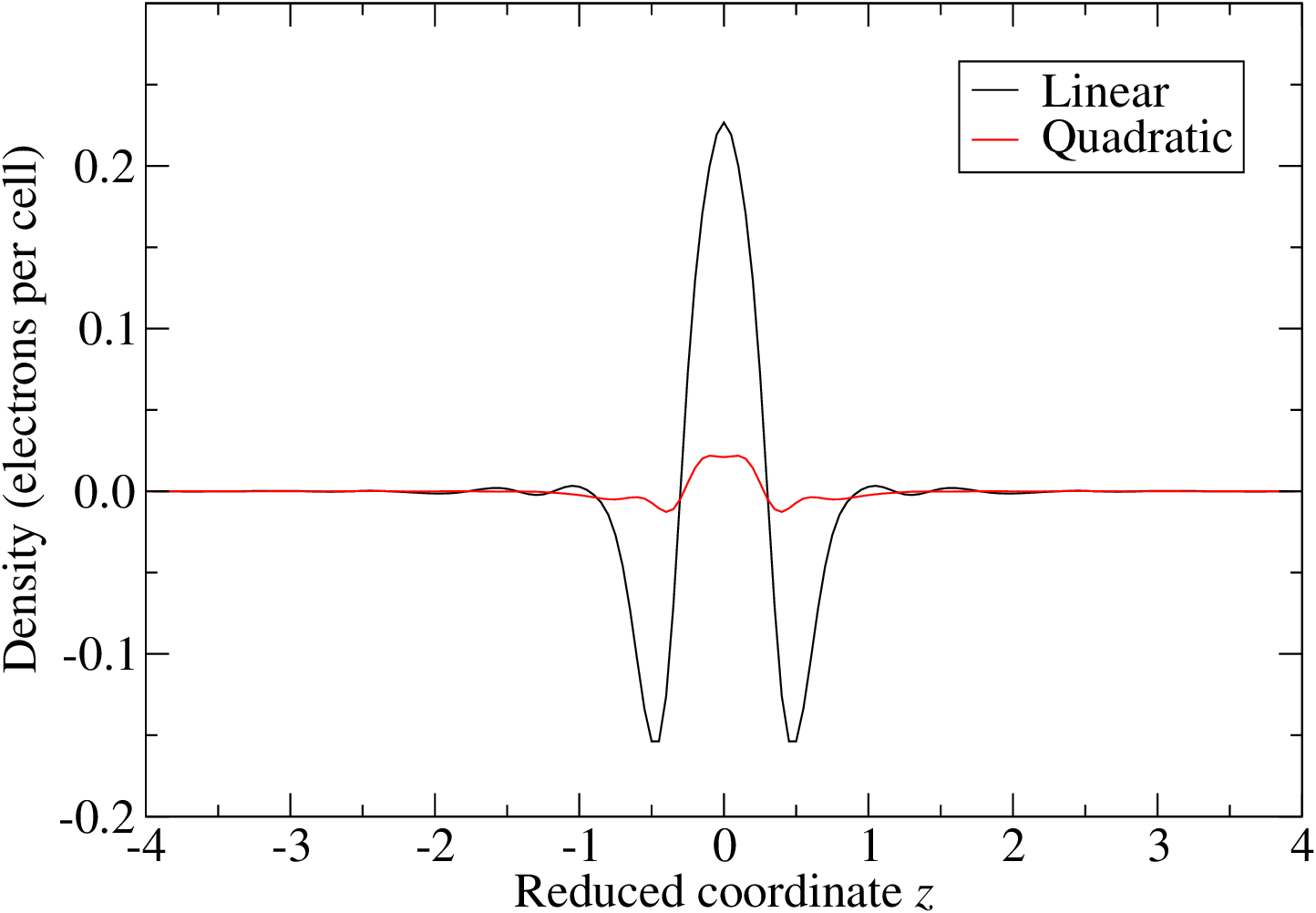}
  \caption{\label{fig:GaAs_mon_den_resp_r} (Color online) Linear and
  quadratic density response to a monatomic Ga perturbation of
  Al$_{0.5}$Ga$_{0.5}$As.}
\end{figure}

The linear and quadratic density response to a monatomic perturbation
in Al$_{0.5}$Ga$_{0.5}$As are shown in Fig.\
\ref{fig:GaAs_mon_den_resp_r}.
The functions plotted here are planar averages of $\Delta
n^{\alpha}_{\vect{R}} (\vect{x})$ and $\Delta
n^{\alpha\alpha}_{\vect{R}\vect{R}} (\vect{x})$ for the case $\alpha =
\mathrm{Ga}$.  Since the supercell used here is the same as that of
the (001) superlattice in Figs.\ \ref{fig:GaAs_slat_overview} and
\ref{fig:GaAs_slat_closeup}, the perturbation actually consists of a
{\em plane} of Ga atoms (or rather an infinitesimal perturbation
toward Ga).  This perturbation has $D_{2d}$ symmetry, so the planar
average shown in Fig.\ \ref{fig:GaAs_mon_den_resp_r} is symmetric.
Note that the monatomic quadratic response is about an order of
magnitude smaller than the linear response.

\begin{figure}
  \includegraphics[width=8.5cm,clip]{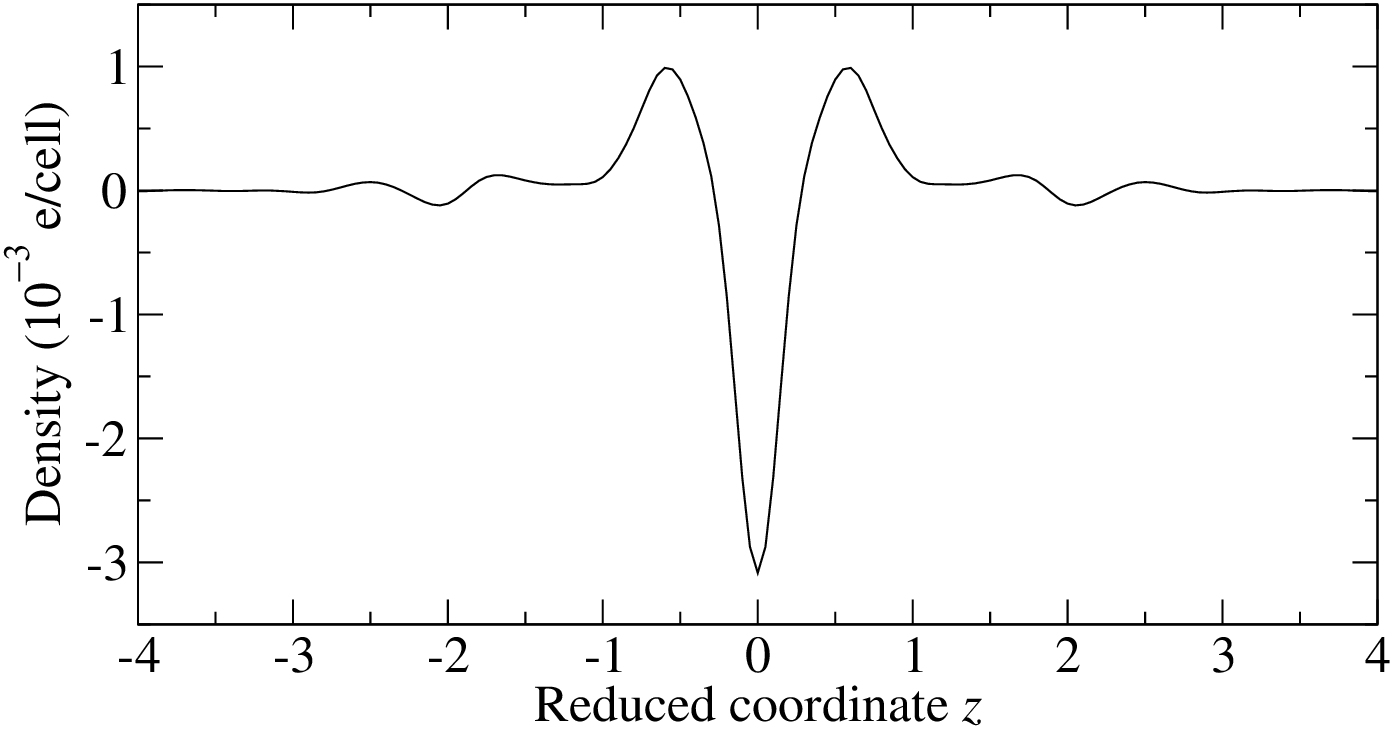}
  \caption{\label{fig:GaAs_dia_den_resp} Quadratic density response to
  a diatomic Ga--Ga perturbation of Al$_{0.5}$Ga$_{0.5}$As.  The
  perturbations are applied at $z = -0.5$ and $z = +0.5$.}
\end{figure}

The quadratic response to a diatomic perturbation in
Al$_{0.5}$Ga$_{0.5}$As is shown in Fig.\ \ref{fig:GaAs_dia_den_resp}.
This perturbation also has $D_{2d}$ symmetry, so its qualitative
features are similar to those of the monatomic response (except that
it is centered on an anion plane rather than a cation plane).  The
perturbation planes in Fig.\ \ref{fig:GaAs_dia_den_resp} are separated
by one monolayer (i.e., $\tfrac12 a$).  The calculations performed in
this paper include diatomic perturbations out to a separation of four
monolayers.  Perturbations beyond this distance were neglected because
they yield contributions of less than 0.1 meV.

The second-nearest-neighbor diatomic response shown in Fig.\
\ref{fig:GaAs_dia_den_resp} is the leading contribution to
$\Gamma_1$--$X_1$ coupling in GaAs/AlAs (001) superlattices.  A
comparison with Fig.\ \ref{fig:GaAs_mon_den_resp_r} shows that this
term is nearly two orders of magnitude smaller than the linear
response, which is the leading contribution to $\Gamma_1$--$X_3$
coupling.  Hence, these results indicate that $\Gamma_1$--$X_1$
coupling is indeed much smaller than $\Gamma_1$--$X_3$ coupling, as
suggested in Refs.\ \onlinecite{WaZu97} and \onlinecite{Fore98b}.

\begin{figure}
  \includegraphics[width=8.5cm,clip]{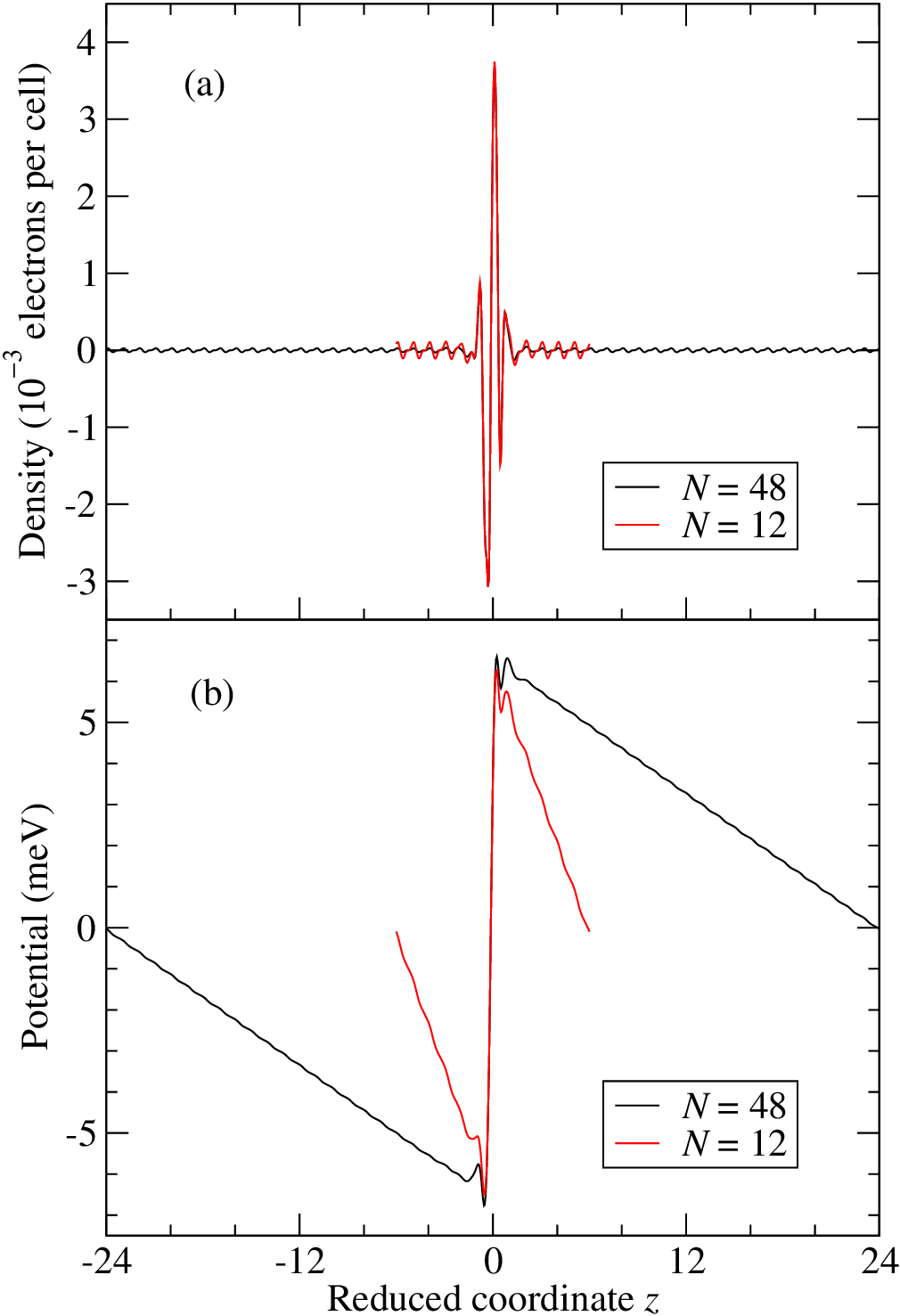}
  \caption{\label{fig:InP_dia_den_resp_r} (Color online) Quadratic
  response to a diatomic As--Ga perturbation of
  In$_{0.765}$Ga$_{0.235}$As$_{0.5}$P$_{0.5}$: (a) electron density,
  (b) Hartree potential.  The perturbations are applied at $z = -0.5$
  (As) and $z = +0.5$ (Ga).  Results are given for supercells with
  periods $N=48$ and $N=12$.}
\end{figure}

Qualitatively different results are obtained for the perturbation of
two different atoms of the In$_{0.765}$Ga$_{0.235}$As$_{0.5}$P$_{0.5}$
reference crystal in Fig.\ \ref{fig:InP_dia_den_resp_r}.
This diatomic perturbation has symmetry $C_{2v}$, so its planar
average is not symmetric.  It therefore has a nonvanishing dipole
moment, which is readily apparent from the figure.  The dipole
generates a rapid change in the Hartree potential, as shown in Fig.\
\ref{fig:InP_dia_den_resp_r}(b).  In order to satisfy the periodic
boundary conditions, this steplike change in potential must be
accompanied by a macroscopic electric field extending over the entire
supercell. \cite{note:pbcfield,Fraser96} This field polarizes the
reference crystal, producing small oscillations in the density and
potential \cite{KuncResta83} that are visible away from $z = 0$.
Figure \ref{fig:InP_dia_den_resp_r} shows calculations performed on
two supercells, one with 48 monolayers (the same as the superlattice
in Figs.\ \ref{fig:InP_slat_overview} and \ref{fig:InP_slat_closeup})
and one with 12 monolayers.  It is evident that the polarization
amplitude is inversely proportional to the period of the supercell.

It is these dipole moments in the diatomic quadratic response that are
responsible for the slight interface asymmetry and macroscopic
electric field referred to in the discussion of Fig.\
\ref{fig:InP_slat_overview}.  Such effects have been studied
previously in Refs.\ \onlinecite{DanZun92} (numerically) and
\onlinecite{Fore05a} (analytically).

\section{Multipole moments and localization}

\label{sec:multipole}

\subsection{Definition of multipole moments}

\label{subsec:multipole_definition}

As discussed in the Introduction, slowly varying envelope functions
probe the density and potential only in a small neighborhood of each
reciprocal-lattice vector $\vect{G}$.  It is therefore convenient to
introduce long-wavelength approximations for the density and potential
response in the form of power-series expansions of $n(\vect{k} +
\vect{G})$ and $v(\vect{k} + \vect{G})$ with respect to $\vect{k}$.
The expansion coefficients can be found numerically by several
methods.  One is a simple polynomial fitting of the discrete fast
Fourier transform (FFT).  Another is to use analytical manipulations
of the FFT to obtain expansion coefficients in the form of multipole
moments.

In a superlattice, the allowed values of $\vect{k} + \vect{G}$ satisfy
$\vect{k} = k \hat{\vect{z}}$ for small $|\vect{k}|$, where
$\hat{\vect{z}}$ is the direction normal to the interface plane.
Hence
\begin{subequations}
\begin{align}
  n(\vect{k} + \vect{G}) & = \frac{1}{N_{\mathrm{F}}} \sum_{\vect{x}}
  n(\vect{x}) e^{-i (\vect{k} + \vect{G}) \cdot \vect{x}} \\ & =
  \sum_{l=0}^{\infty} (-ik)^l n_l (\vect{G}) , \label{eq:power}
\end{align}
\end{subequations}
where $\vect{x}$ is a coordinate on the FFT grid, $N_{\mathrm{F}}$
is the number of such points, and the expansion coefficients are
\begin{subequations} \label{eq:multipole}
\begin{align}
  n_l(\vect{G}) & = \frac{1}{l! N_{\mathrm{F}}} \sum_{\vect{x}} z^l
  n(\vect{x}) e^{-i \vect{G} \cdot \vect{x}} \\ & = \frac{1}{l!}
  \sum_{\vect{k}} [\zeta_l(\vect{k} - \vect{G})]^* n(\vect{k}) ,
\end{align}
\end{subequations}
where $\zeta_l(\vect{k})$ is the FFT of $z^l$.  The coefficient
$n_l(\vect{G})$ is, to within a factor of $l!$, just the multipole
moment \cite{BaReBaPe89,Jackson99} of order $2^l$ of the function
$n(\vect{x}) e^{-i \vect{G} \cdot \vect{x}}$.

\subsection{Order of terms included}

\label{subsec:multipole_order}

For small $k$, the power series (\ref{eq:power}) can be approximated
accurately by including only a few terms with small $l$.  This paper
adopts the approximation scheme defined in Refs.\ \onlinecite{Fore05a}
and \onlinecite{Fore05b}, in which the power series for the linear and
quadratic potential response $v^{(1)}(\vect{k} + \vect{G})$ and
$v^{(2)}(\vect{k} + \vect{G})$ are terminated at the upper limits
\begin{subequations} \label{eq:lmax}
\begin{equation}
  l_v^{(1)} = 2 , \qquad l_v^{(2)} = 0 . \label{eq:lvmax}
\end{equation}
These power series can be used for the pseudopotential and
exchange-correlation potential response, which are analytic functions
of $\vect{k}$.  However, the Hartree potential (\ref{eq:V_Har}) is
nonanalytic, so the expansion (\ref{eq:power}) must be applied to the
density $n(\vect{k} + \vect{G})$ instead.  To obtain an accuracy
equivalent to (\ref{eq:lvmax}), the upper limits for the density power
series are extended when $G = 0$:
\begin{equation}
  l_n^{(1,2)} (\vect{G}) = l_v^{(1,2)} + 2 \delta_{\vect{G}\vect{0}} ,
  \label{eq:lnmax}
\end{equation}
\end{subequations}
because of the $k^{-2}$ factor in the Hartree potential
(\ref{eq:V_Har}).

When $G \ne 0$, the Hartree potential is an analytic function of
$\vect{k}$ (for small $k$), and the power series expansion of $|
\vect{k} + \vect{G} |^{-2}$ generates the following Hartree
corrections to the analytic potential coefficients:
%% \begin{subequations}
%% \begin{align}
%%   \Delta v_{0}^{(1,2)} (\vect{G}) & = \frac{4\pi}{G^2} n_{0}^{(1,2)}
%%   (\vect{G}) ,
%%   \label{eq:dv0} \\
%%   \Delta v_{1}^{(1)} (\vect{G}) & = \frac{4\pi}{G^2} \biggl[
%%   n_{1}^{(1)} (\vect{G}) - \frac{i 2 G_z}{G^2} n_{0}^{(1)} (\vect{G})
%%   \biggr] , \\
%%     \Delta v_{2}^{(1)} (\vect{G}) & = \frac{4\pi}{G^2} \biggl[
%%       n_{2}^{(1)} (\vect{G}) - \frac{i 2 G_z}{G^2} n_{1}^{(1)}
%%       (\vect{G}) + \biggl( \frac{1}{G^2} - \frac{4 G_z^2}{G^4} \biggr)
%%       n_{0}^{(1)} (\vect{G}) \biggr] .
%% \end{align}
%% \end{subequations}
\begin{subequations}
  \begin{align}
    \Delta v_{0}^{(1,2)} (\vect{G}) & = \frac{4\pi}{G^2} n_{0}^{(1,2)}
    (\vect{G}) ,
    \label{eq:dv0} \\
    \Delta v_{1}^{(1)} (\vect{G}) & = \frac{4\pi}{G^2} \biggl[
      n_{1}^{(1)} (\vect{G}) - \frac{i 2 G_z}{G^2} n_{0}^{(1)} (\vect{G})
      \biggr] , \\
    \begin{split}
      \Delta v_{2}^{(1)} (\vect{G}) & = \frac{4\pi}{G^2} \biggl[
	n_{2}^{(1)} (\vect{G}) - \frac{i 2 G_z}{G^2} n_{1}^{(1)}
	(\vect{G}) \\ & \qquad \qquad {} + \biggl( \frac{1}{G^2} -
	\frac{4 G_z^2}{G^4} \biggr) n_{0}^{(1)} (\vect{G}) \biggr] .
    \end{split}
  \end{align}
\end{subequations}
When $G = 0$, analytic potential contributions are also obtained from
the terms
\begin{equation}
  \Delta v_l(\vect{0}) = -4 \pi n_{l+2} (\vect{0}) \qquad (l \ge 0),
  \label{eq:addHart0}
\end{equation}
where $0 \le l \le 2$ for the linear response and $l = 0$ for the
quadratic response.  In a superlattice, the only nonanalytic terms
arise from the density monopole, dipole, and quadrupole coefficients
at $G = 0$.  The quadrupole term is given by $4 \pi n_2(\vect{0})
\delta_{\vect{k}\vect{0}}$, which merely contributes an overall
constant shift in potential.  The monopole term is zero for the
isovalent perturbations considered here, while the linear dipole term
vanishes for perturbations with $T_d$ or $D_{2d}$ symmetry.  The
quadratic dipole term is, however, nonzero for diatomic perturbations
with $C_{2v}$ symmetry, as shown above in Fig.\
\ref{fig:InP_dia_den_resp_r}.

\subsection{Localization}

\label{subsec:multipole_localization}

Since the FFT grid is finite, the multipole moment $n_l(\vect{G})$
always has a well defined value.  Nevertheless, if the response
$n(\vect{x})$ is not well localized, the value of $n_l(\vect{G})$,
although finite, does not converge to a meaningful value in the limit
of large supercells.  It is therefore necessary to check in each case
whether the localization is sufficient to use Eq.\
(\ref{eq:multipole}) for the desired values of $l$.

The localization of the response is not merely a numerical detail; it
is fundamental to the physics of the insulating
state. \cite{Kohn64,VaKiSm93} Phenomenological envelope-function
models typically assume that the heterostructure Hamiltonian can be
expressed as a power series in $\vect{k}$, but the validity of such an
assumption depends crucially on the localization of the response in an
insulator. \cite{Fore05a,Fore05b} If the response is not localized,
the entire foundation for the power-series expansion of the
Hamiltonian breaks down.  Establishing that the response is localized
is therefore a key element in the justification of the standard
phenomenological approach to envelope-function theory.  Numerical
error inevitably masks the underlying localization at some point, but
it is important to ensure that this error is insignificant for all
practical purposes.

\begin{figure}
  \includegraphics[width=8.5cm,clip]{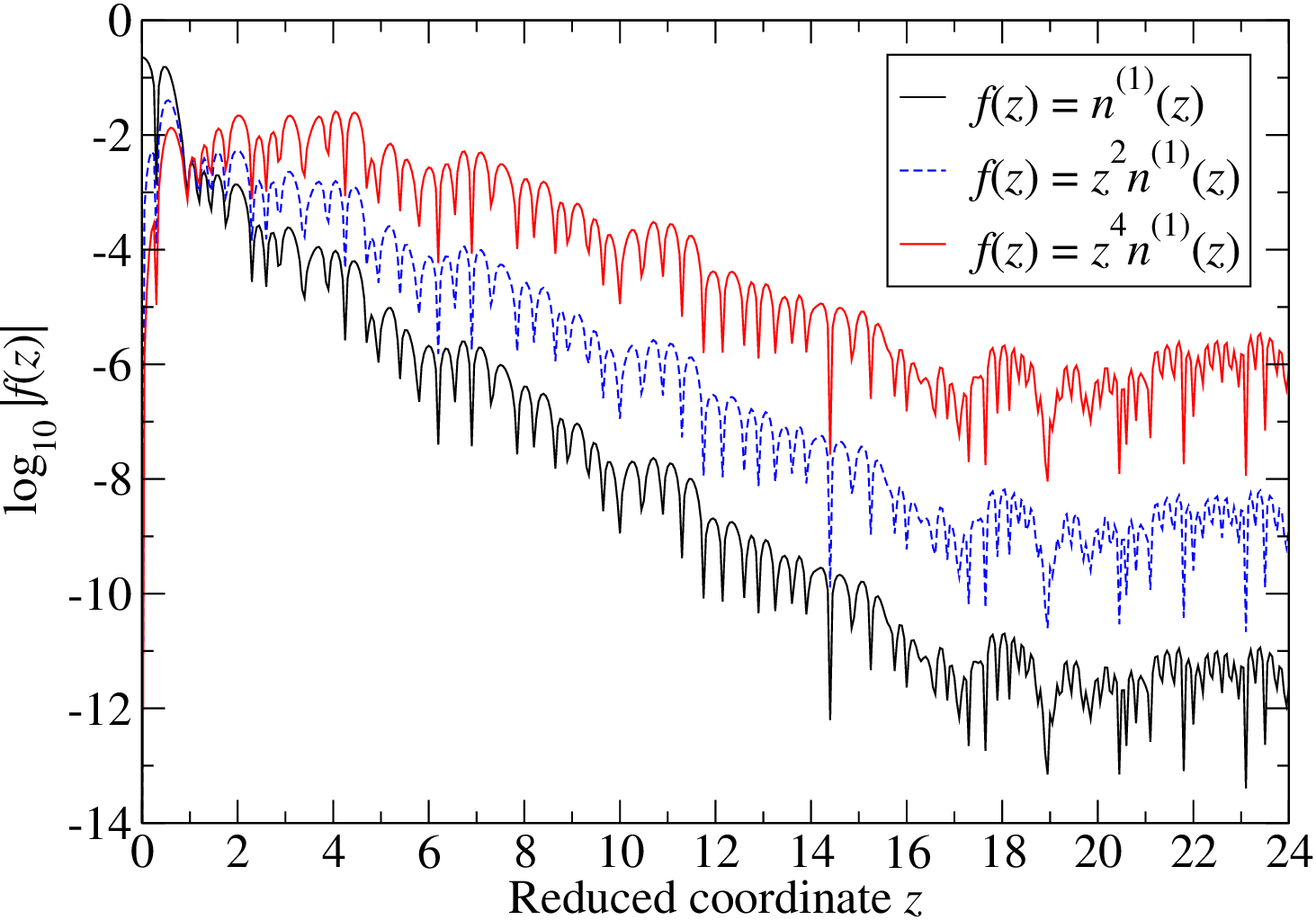}
  \caption{\label{fig:GaAs_log_mon_den} (Color online) Log-scale plot
  of linear density response in Fig.\ \ref{fig:GaAs_mon_den_resp_r}.}
\end{figure}

This is done in Fig.\ \ref{fig:GaAs_log_mon_den} for the linear
density response of Al$_{0.5}$Ga$_{0.5}$As in the cases $l = 0$, 2,
and 4.
The response is quasi-exponentially localized over nearly 10 orders of
magnitude, to a distance of about 17 monolayers from the perturbation.
At this point it flattens out due to the nonanalyticity of the
potential-energy smoothing function in Fig.\
\ref{fig:GaAs_kinetic_divisor}.  If the potential energy is not
smoothed, the quasi-exponential localization extends for another order
of magnitude, at which point numerical error takes over.  However, the
localization obtained here is clearly sufficient to calculate accurate
values of the linear quadrupole and hexadecapole moments.  On the
other hand, if the plane-wave cutoff in the kinetic energy is not
smoothed, $z^4 n^{(1)}(z)$ is not adequately localized, and the
hexadecapole moment ($l = 4$) has a nonsensical value for the
supercell considered here.

\begin{figure}
  \includegraphics[width=8.5cm,clip]{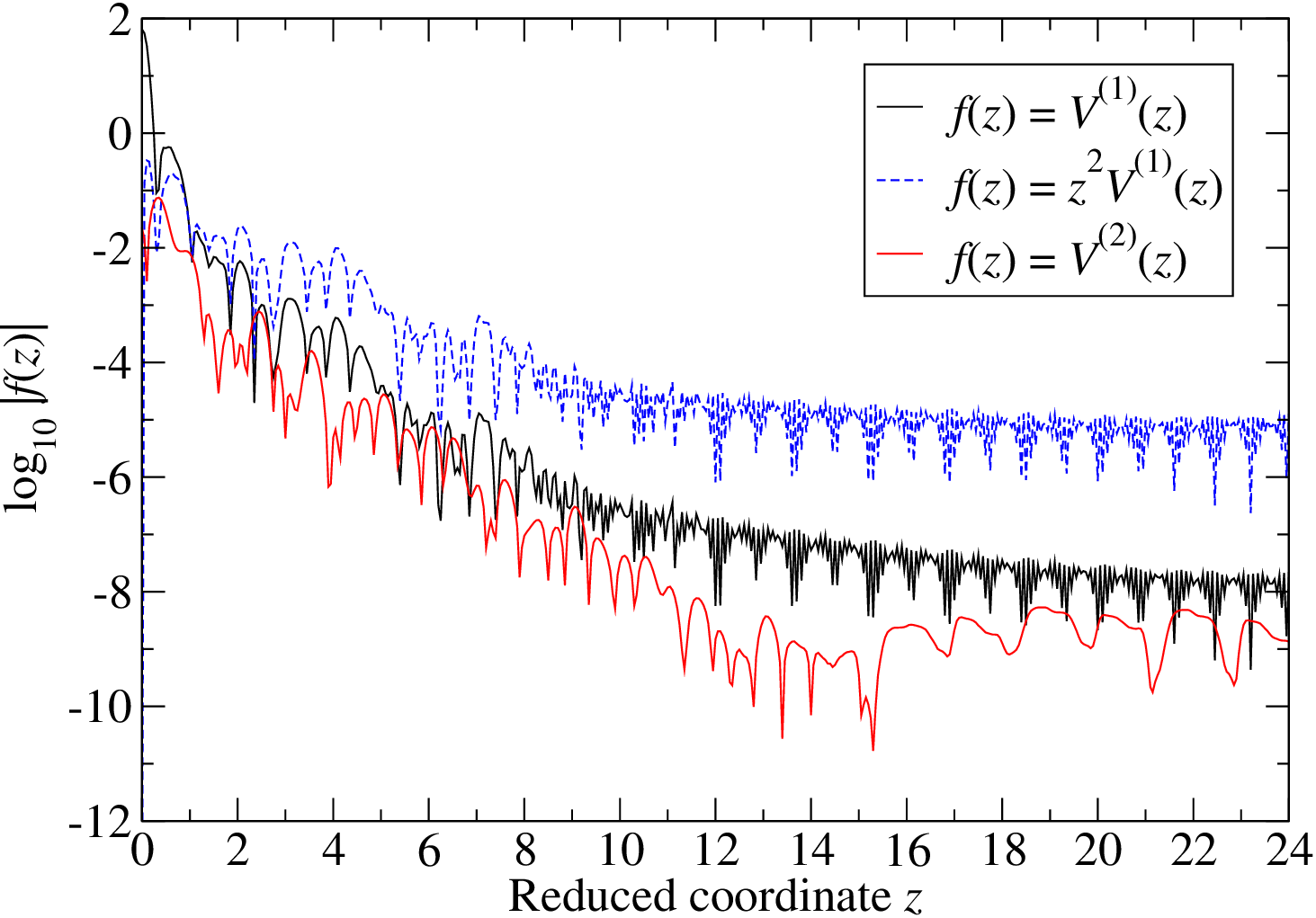}
  \caption{\label{fig:GaAs_log_mon_pot} (Color online) Log-scale plot
  of linear and quadratic pseudopotential plus exchange-correlation
  potential for a monatomic Ga perturbation of
  Al$_{0.5}$Ga$_{0.5}$As.}
\end{figure}

The linear and quadratic terms in those parts of the local potential
response that are expected to be localized (i.e., the ionic
pseudopotential and the exchange-correlation potential) are shown in
Fig.\ \ref{fig:GaAs_log_mon_pot}.
Here the features seen in the linear response for $|z| \gtrsim 10$ are
the remnants of Gibbs oscillations in the local ionic pseudopotential
that remain even after smoothing of the plane-wave cutoff.  (These
Gibbs oscillations are not seen in the quadratic response because the
ionic pseudopotential is purely linear.)  The localization here is not
quite as good as for the density, but it is still sufficient to
calculate the desired multipole moments.

\begin{figure}
  \includegraphics[width=8.5cm,clip]{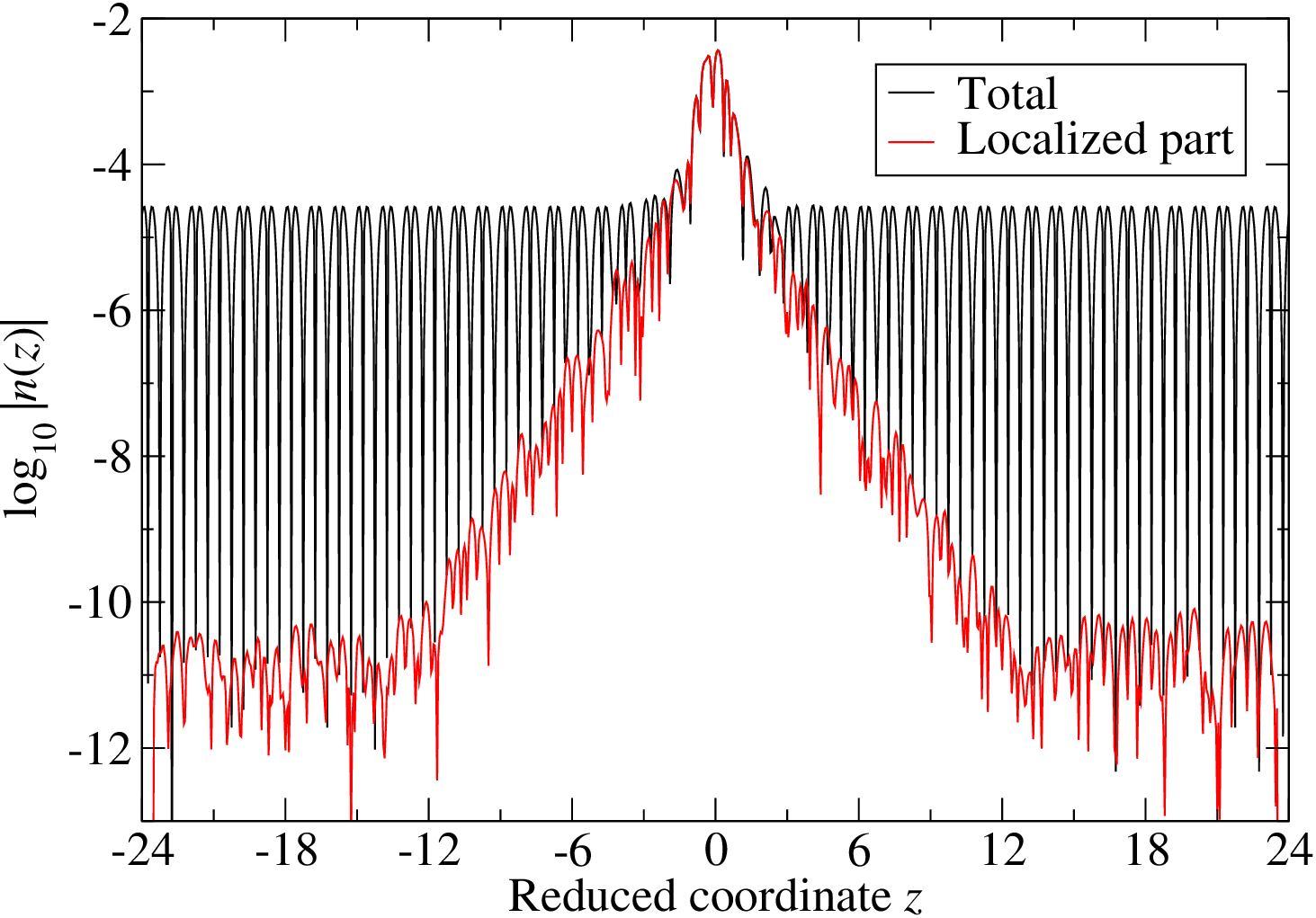}
  \caption{\label{fig:InP_log_dia_den} (Color online) Log-scale plot
  of the diatomic density response in Fig.\
  \ref{fig:InP_dia_den_resp_r}(a).  Both the total density and its
  localized part (excluding the periodic part) are shown.}
\end{figure}

The log-scale plot in Fig.\ \ref{fig:InP_log_dia_den} of the quadratic
density response for $C_{2v}$ diatomic perturbations has additional
features of interest.
The response is not well localized, but this is a physical effect
arising from the periodic oscillations shown in Fig.\
\ref{fig:InP_dia_den_resp_r}.  The periodic part can be separated from
the rest by evaluating it in the unit cell farthest from the origin.
Subtracting the periodic part from the total leaves a remainder that
should be localized.  This is confirmed in Fig.\
\ref{fig:InP_log_dia_den}, which shows both the total quadratic
response and the localized remainder.

The localized part of the diatomic response can be used to obtain
multipole moments as above.  The periodic part is even simpler,
because it merely modifies the values of coefficients in the bulk
Hamiltonian (as discussed in greater detail below).

\subsection{Calculated multipole moments}

\begin{figure}
  \includegraphics[width=8.5cm,clip]{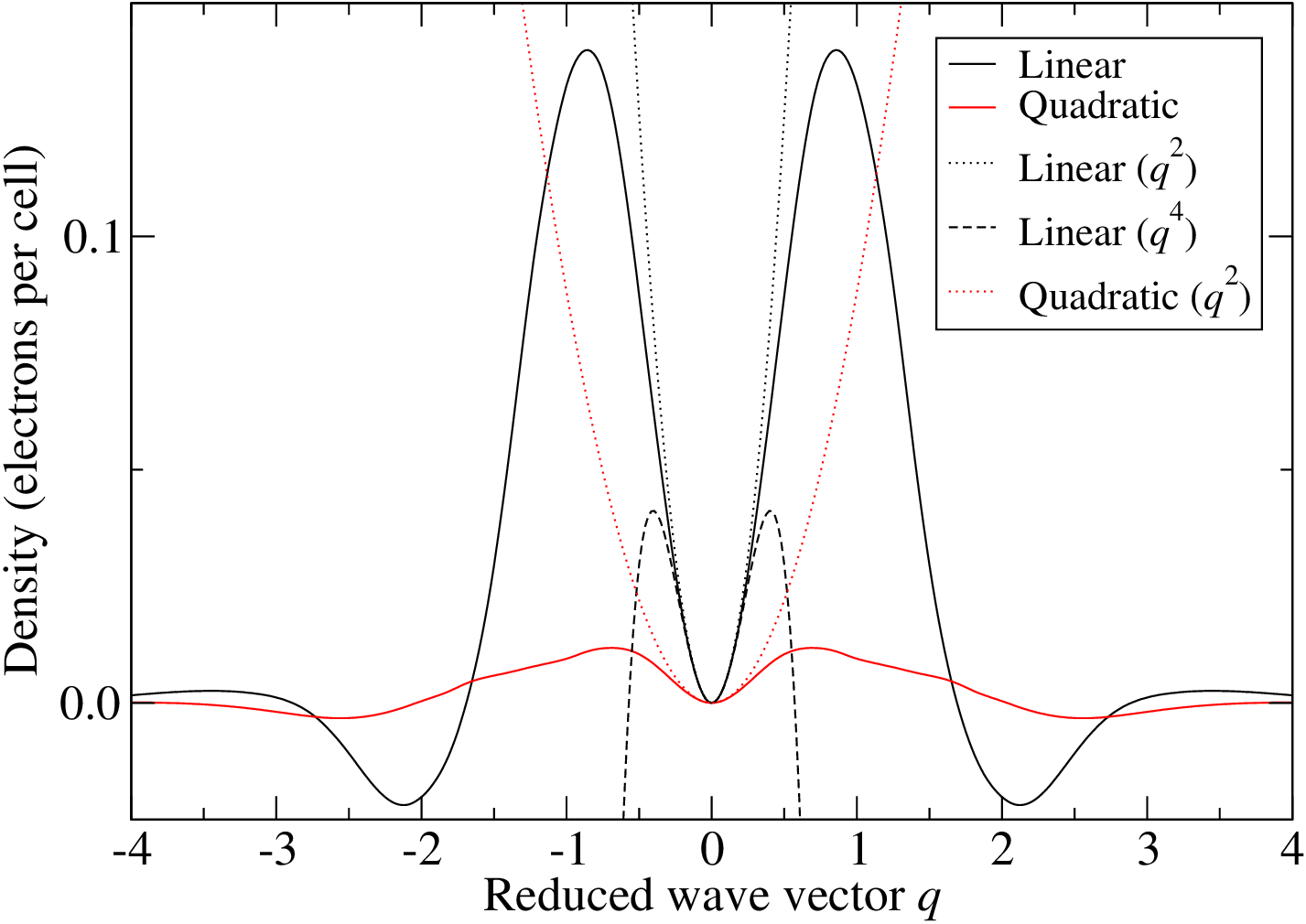}
  \caption{\label{fig:GaAs_mon_den_resp_k} (Color online) Linear and
  quadratic density response to a monatomic Ga perturbation of
  Al$_{0.5}$Ga$_{0.5}$As.  The solid lines are Fourier transforms of
  the functions in Fig.\ \ref{fig:GaAs_mon_den_resp_r}.  The dotted
  and dashed lines are quadratic and quartic polynomials defined by
  the quadrupole and hexadecapole moments.}
\end{figure}

\begin{figure}
  \includegraphics[width=8.5cm,clip]{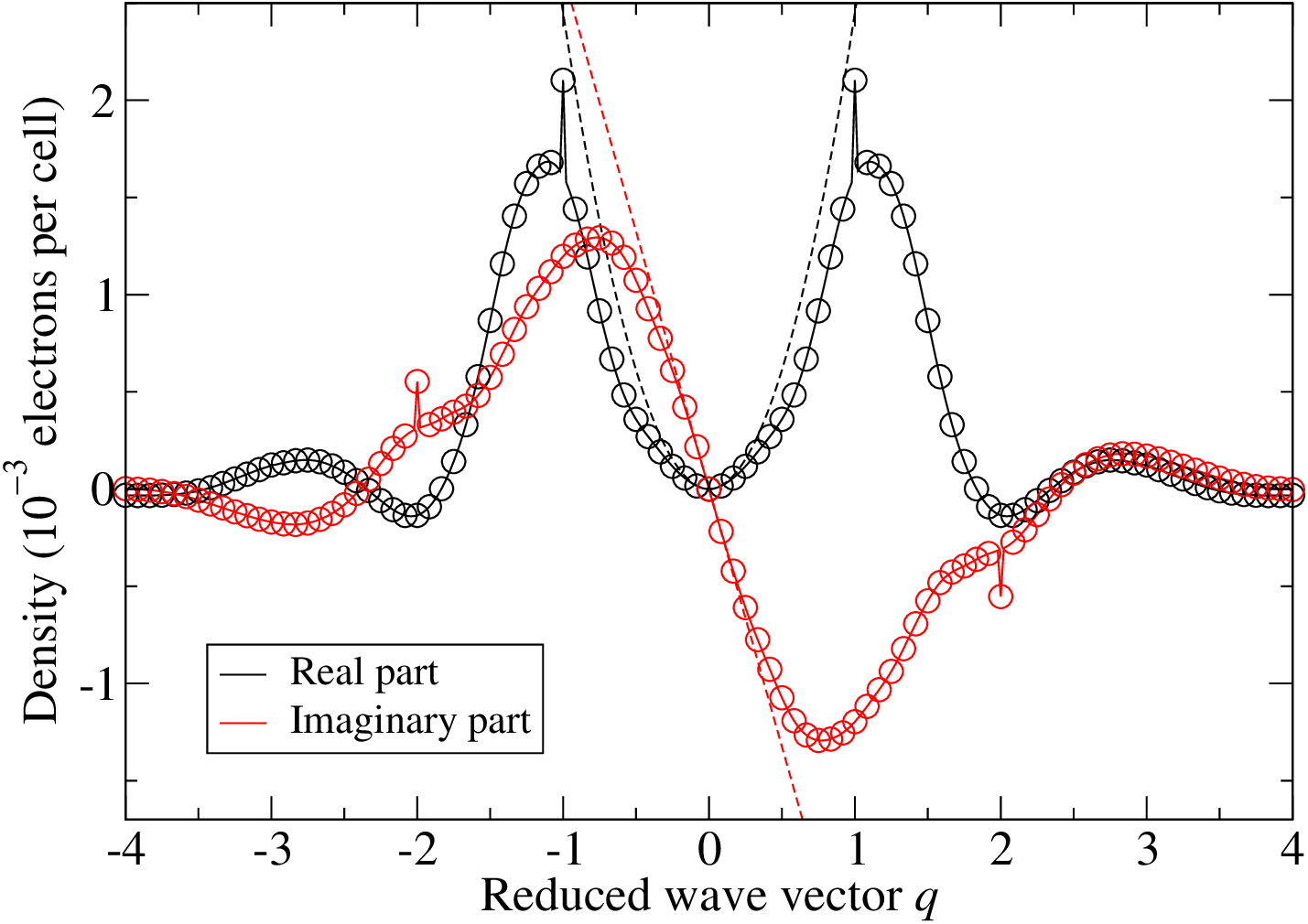}
  \caption{\label{fig:InP_dia_den_resp_k} (Color online) Fourier
  transform of the diatomic density response in Fig.\
  \ref{fig:InP_dia_den_resp_r}(a).  Solid lines are for a supercell
  with $N = 48$ monolayers, while circles are for $N = 12$.  The FFT
  is scaled by $N$ so that the results are comparable.  Dashed lines
  are quadratic (real) and linear (imaginary) polynomials
  corresponding to the calculated quadrupole and dipole moments.}
\end{figure}

The results obtained from the calculated density multipole moments
$n_l(\vect{G} = \vect{0})$ are shown in Figs.\
\ref{fig:GaAs_mon_den_resp_k} and \ref{fig:InP_dia_den_resp_k} for
Al$_{0.5}$Ga$_{0.5}$As and
In$_{0.765}$Ga$_{0.235}$As$_{0.5}$P$_{0.5}$, respectively.
The truncated power series expansions for the $\vect{k}$-space density
response are clearly valid only for small wave vectors.  For the
$C_{2v}$ perturbation in Fig.\ \ref{fig:InP_dia_den_resp_k}, the
reduced symmetry generates a nonvanishing imaginary part in the FFT.
This part has a nonzero slope at the origin, leading to the dipole
moment observed previously in Fig.\ \ref{fig:InP_dia_den_resp_r}.

There are also noticeable spikes in the FFT at nonzero
reciprocal-lattice vectors.  These correspond to the periodic part of
the response discussed above (Sec.\
\ref{subsec:multipole_localization}), whose physical origin is the
polarization of the reference crystal by the macroscopic electric
field \cite{note:pbcfield} in Fig.\ \ref{fig:InP_dia_den_resp_r}.  The
amplitude of the spikes is obtained most accurately by the method
described above, in which the periodic part of the quadratic response
is evaluated in coordinate space in the unit cell farthest from the
origin and then Fourier transformed.  The resulting Fourier series
coefficients occur only at the bulk reciprocal-lattice vectors
$\vect{G}$, which means that their contribution to the
envelope-function Hamiltonian \cite{Fore07b} has the same formal
structure as the bulk \kp\ Hamiltonian of the reference crystal.
However, these bulk polarization terms have the $C_{2v}$ symmetry of
the dipole, not the $T_{d}$ symmetry of the reference crystal.  As
shown in Ref.\ \onlinecite{Fore07b}, the net effect of the $C_{2v}$
dipoles and their associated bulk polarization is to contribute about
one third of the total splitting of the quasidegenerate $X$ and $Y$
valence states (which would be degenerate under $D_{2d}$ symmetry) in
In$_{0.53}$Ga$_{0.47}$As/InP superlattices.

\section{Error in the linear and quadratic approximations}

\label{sec:quaderr}

\subsection{Electron density and Hartree potential}

\label{subsec:quaderr_den}

To test the validity of truncating the power series in the nonlinear
response (\ref{eq:nonlinear}), the effects of various truncations were
examined by comparing them with the exact superlattice density and
potential given previously in Figs.\ \ref{fig:GaAs_slat_overview},
\ref{fig:GaAs_slat_closeup}, \ref{fig:InP_slat_overview}, and
\ref{fig:InP_slat_closeup}.  The approximations are generally good
enough that they are difficult to distinguish visually from the exact
results.  Therefore, only the error in the approximations is plotted
here.

\begin{figure}
  \includegraphics[width=8.5cm,clip]{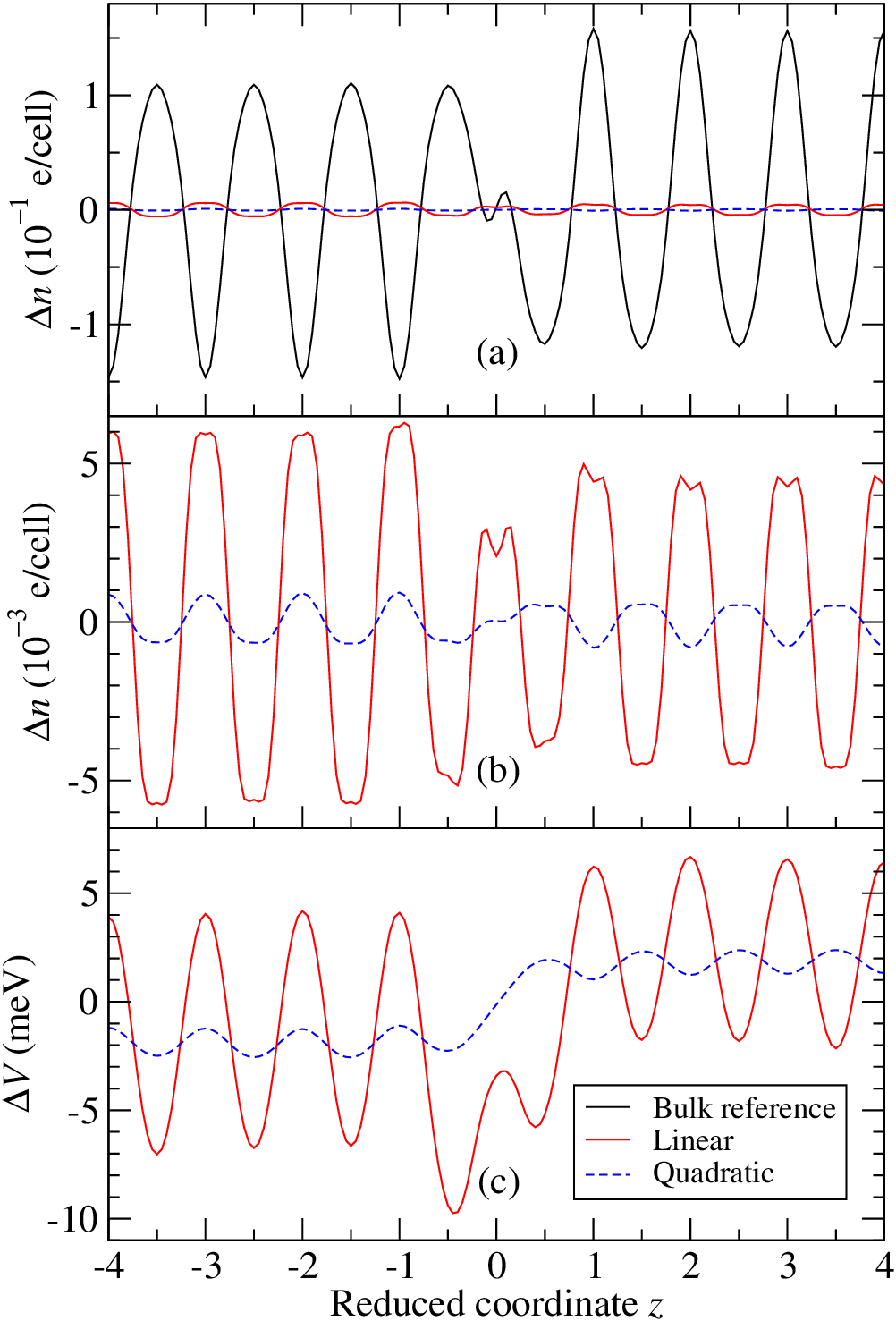}
  \caption{\label{fig:GaAs_slat_den_err} (Color online) Errors in the
  electron density and Hartree potential of a GaAs/AlAs superlattice
  under various approximations: (a)~Errors in the reference crystal
  density and the densities obtained in the linear and quadratic
  approximations.  (b)~Expanded view of the linear and quadratic
  errors in (a).  (c)~Errors in the linear and quadratic
  approximations to the Hartree potential.}
\end{figure}

The results for the GaAs/AlAs superlattice are shown in Fig.\
\ref{fig:GaAs_slat_den_err}.
Part (a) shows the error when the superlattice density is approximated
by the periodic reference crystal density, comparing it with the error
in the linear and quadratic approximations.  The latter are both small
on the scale of the reference crystal error (which is itself small on
the scale of the reference density), so the linear and quadratic
errors are plotted on an expanded scale in part (b).  The
corresponding errors in the Hartree potential are shown in part (c).
It can be seen that the quadratic approximation is quite accurate,
with an error of only about $\pm 2$ meV\@.

\begin{figure}
  \includegraphics[width=8.5cm,clip]{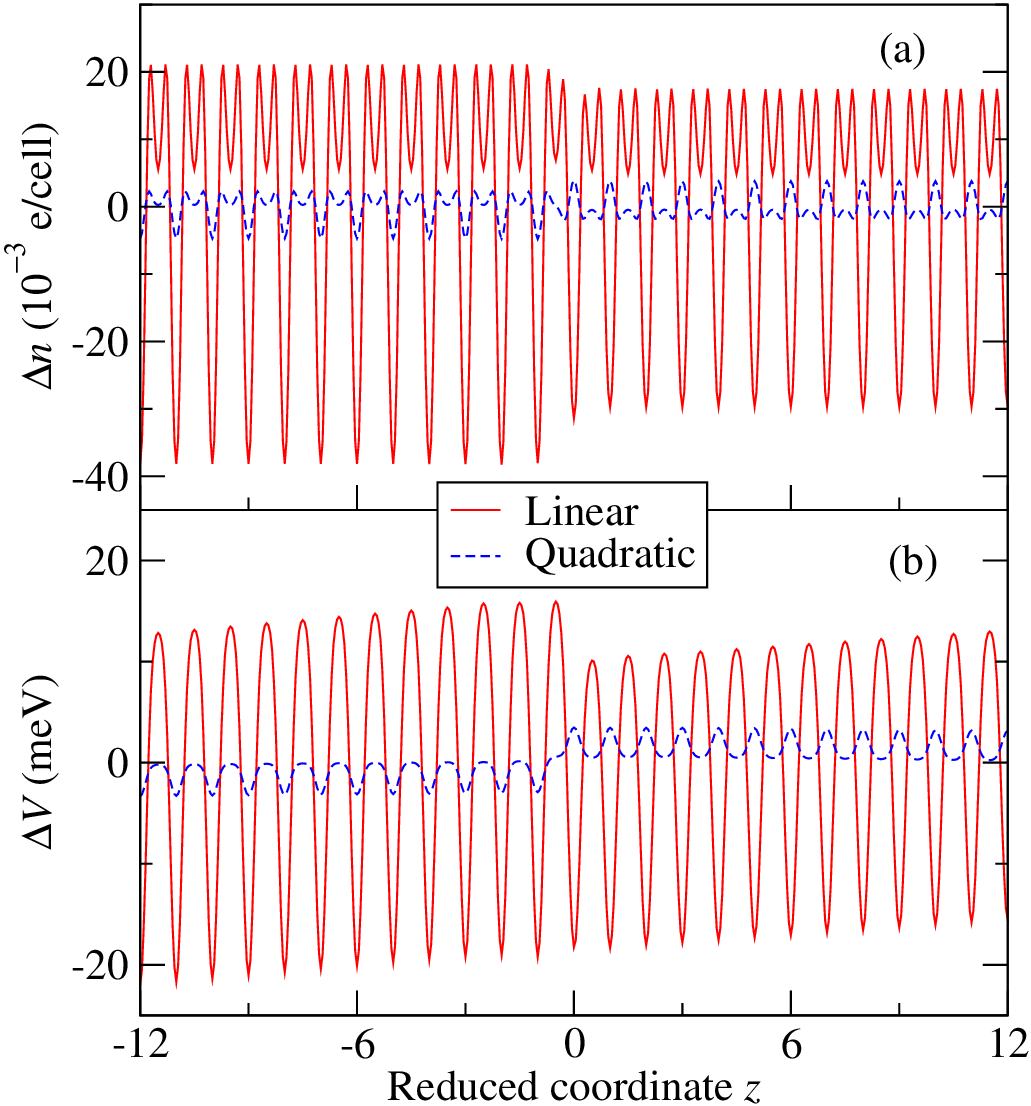}
  \caption{\label{fig:InP_slat_den_err} (Color online) Errors in the
  electron density and Hartree potential of an
  In$_{0.53}$Ga$_{0.47}$As/InP superlattice under various
  approximations: (a)~Errors in the linear and quadratic densities.
  (b)~Errors in the linear and quadratic Hartree potentials.}
\end{figure}

The errors in the linear and quadratic approximations for the
In$_{0.53}$Ga$_{0.47}$As/InP superlattice are shown in Fig.\
\ref{fig:InP_slat_den_err}.
Here the error in the quadratic Hartree potential has an oscillation
with an amplitude of about 3 meV superimposed on a shift of about $\pm
1.5$ meV\@.

\subsection{Truncation of multipole expansions}

\label{subsec:trunc_multipole}

\begin{figure}
  \includegraphics[width=8.5cm,clip]{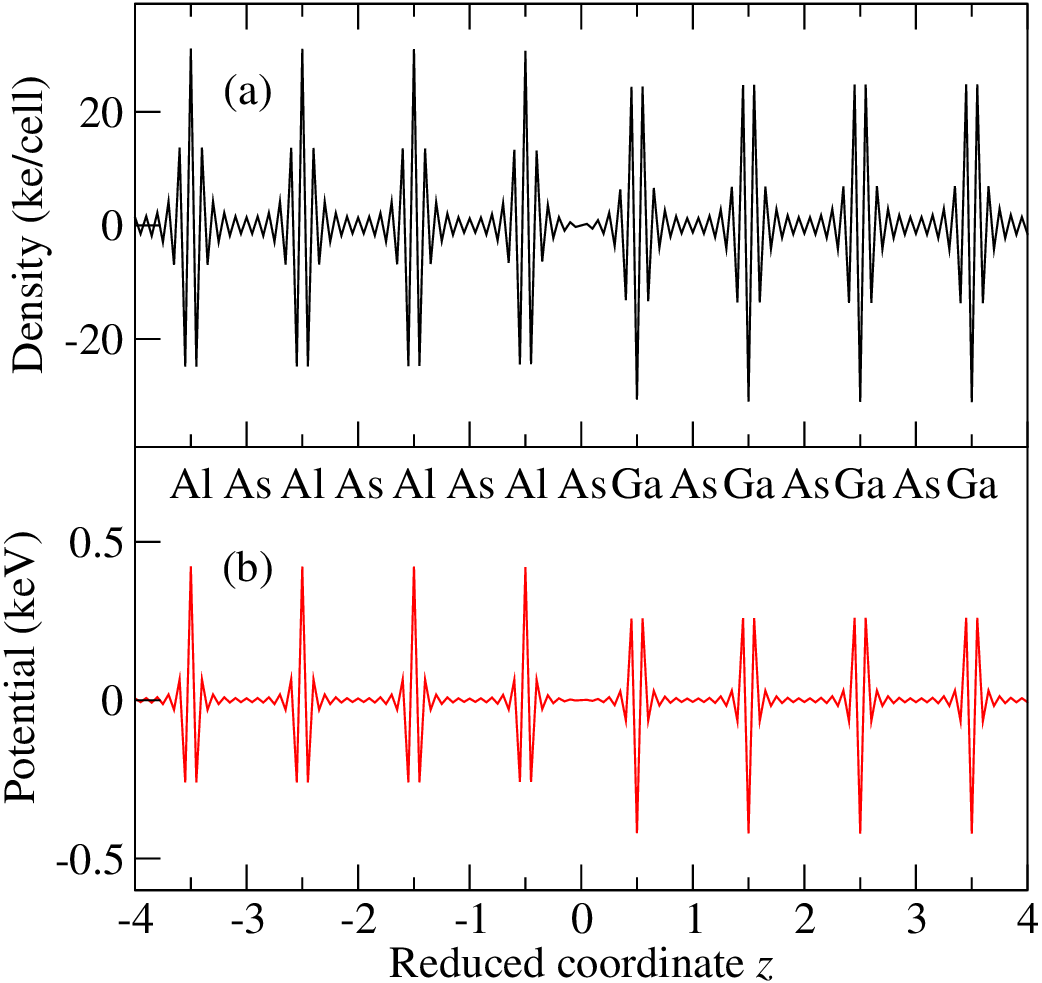}
  \caption{\label{fig:GaAs_slat_den_fit} (Color online) Sum of the
  linear and quadratic perturbations for (a) the electron density and
  (b) the Hartree potential constructed from quadrupole and
  hexadecapole moments in a GaAs/AlAs superlattice.  The perturbations
  are finite only because a discrete coordinate grid is used.  The
  hexadecapole terms are dominant here, although this is no longer
  true after macroscopic averaging.}
\end{figure}

Using multipole expansions for the density and potential introduces an
additional source of error.  To illustrate this, Fig.\
\ref{fig:GaAs_slat_den_fit} shows the electron density and Hartree
potential in coordinate space calculated directly from the FFT of the
truncated multipole expansion (with $0 \le l \le 4$ for the linear
density and $0 \le l \le 2$ for the quadratic density).
Note that the scale of this figure is much larger than any of the
others.  In fact, in a continuous coordinate space, these quantities
would be derivatives of Dirac $\delta$ functions.  Clearly these
multipole expansions yield a very poor approximation of the original
density and potential.

\begin{figure}
  \includegraphics[width=8.5cm,clip]{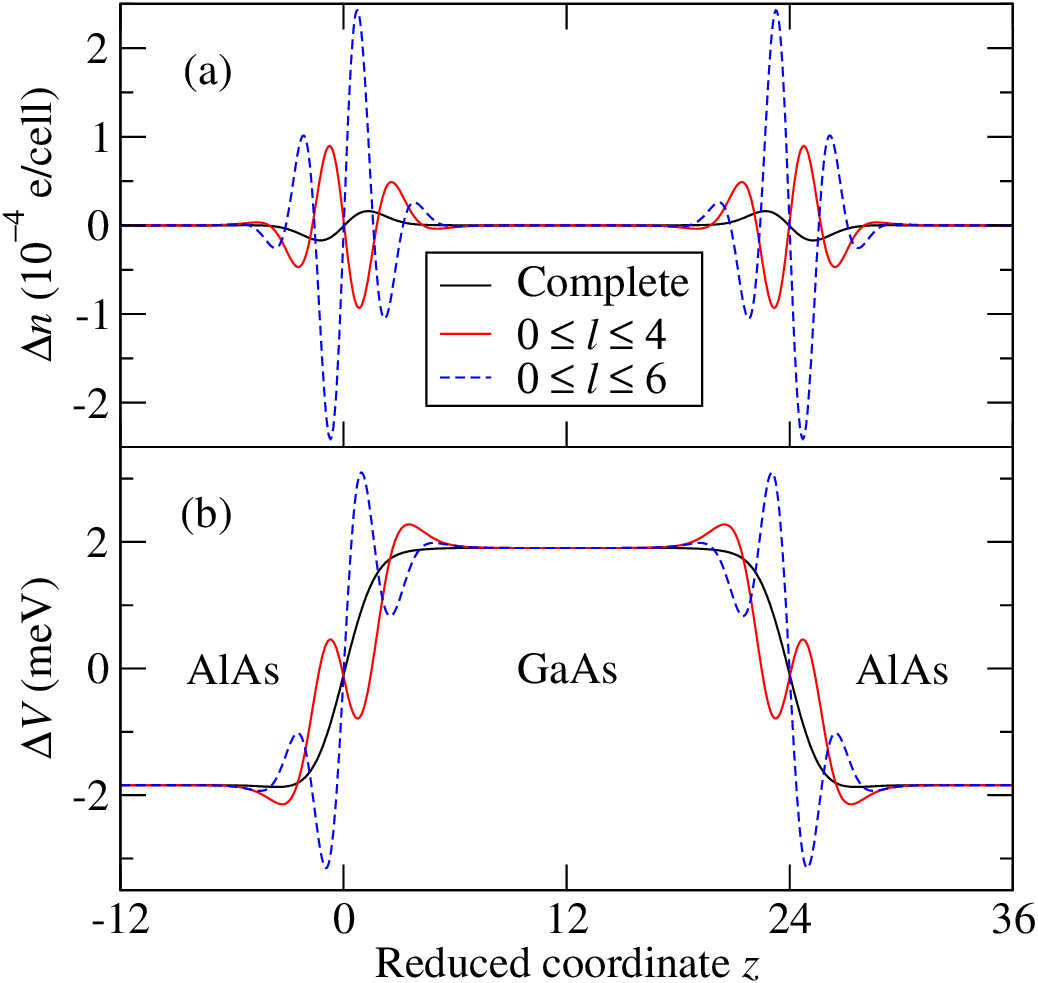}
  \caption{\label{fig:GaAs_slat_fit_avg} (Color online) Errors in the
  macroscopic (a) electron density and (b) Hartree potential of a
  GaAs/AlAs superlattice under various approximations.  The functions
  shown are the errors in the complete quadratic approximation to the
  macroscopic density and potential (i.e., the macroscopic averages of
  the quadratic functions in Fig.\ \ref{fig:GaAs_slat_den_err} with
  $\sigma = 3d$) and the corresponding errors when the linear and
  quadratic responses are generated from truncated multipole
  expansions.  The labels $0 \le l \le 4$ and $0 \le l \le 6$ refer to
  the power series for the linear density; for the quadratic density,
  the corresponding limits are $0 \le l \le 2$ and $0 \le l \le 4$,
  respectively.}
\end{figure}

However, in an envelope-function model, accuracy is only required at
small wave vectors, so it is more appropriate to compare the
macroscopic average of the multipole expansions with the macroscopic
average of the exact results.  This is done in Fig.\
\ref{fig:GaAs_slat_fit_avg}, which shows the errors in the macroscopic
density and Hartree potential for a GaAs/AlAs superlattice.
Here the macroscopic averages were calculated using $\sigma = 3d$ in
Eq.\ (\ref{eq:macroave}), which limits momentum transfers to roughly
$\frac13$ of the bulk $\Gamma$--$X$ distance. \cite{note:macroave}
Such a limit is appropriate for the slowly varying envelope functions
considered in the following paper. \cite{Fore07b} The macroscopic
errors are virtually identical in bulk, but the truncated multipole
approximation introduces some additional error near the interfaces.
This interface error is negligible for the case $\sigma = 3d$ shown
here, although it becomes arbitrarily large in the limit of small
$\sigma$.  This merely reflects the fact that a truncated multipole
expansion is valid only for slowly varying envelopes.

\begin{figure}
  \includegraphics[width=8.5cm,clip]{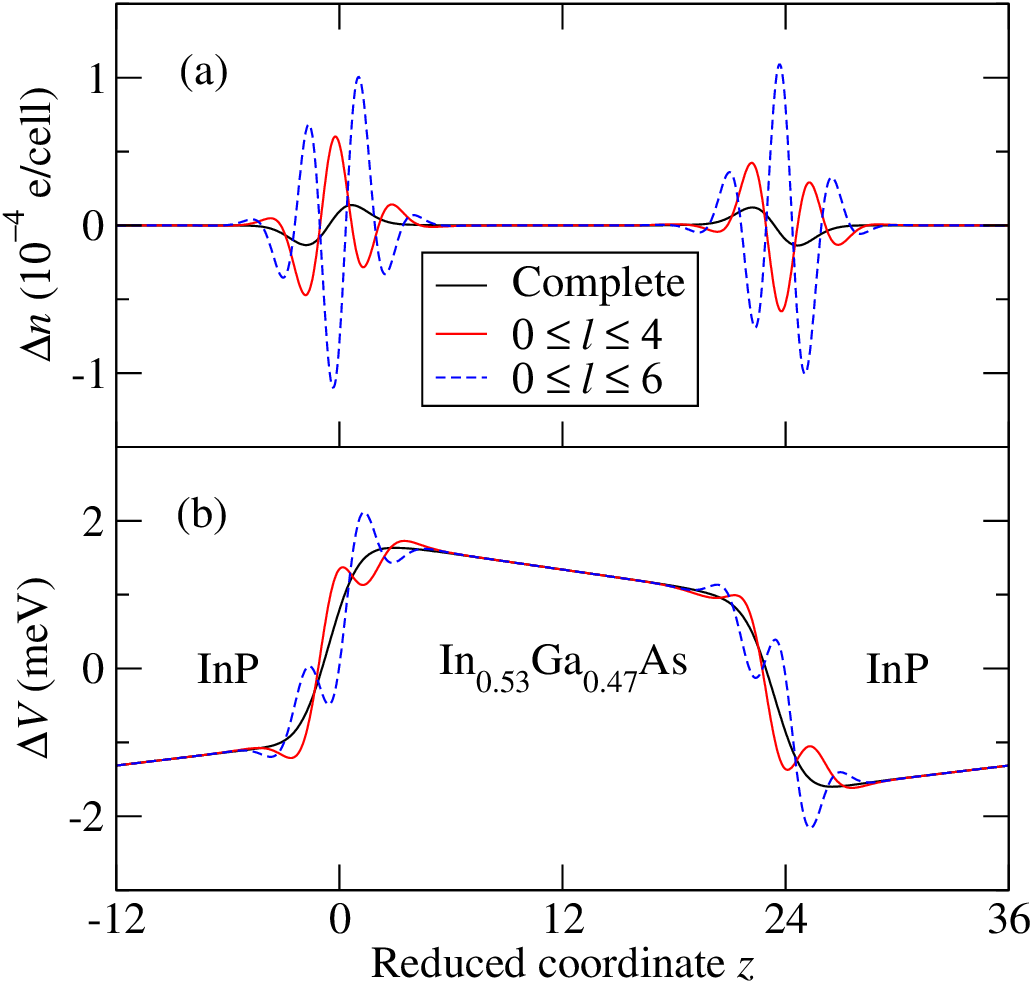}
  \caption{\label{fig:InP_slat_fit_avg} (Color online) Errors in the
  macroscopic (a) electron density and (b) Hartree potential of an
  InP/In$_{0.53}$Ga$_{0.47}$As superlattice: comparison of complete
  quadratic errors with truncated multipole errors, as in Fig.\
  \ref{fig:GaAs_slat_fit_avg}.}
\end{figure}

The corresponding results for In$_{0.53}$Ga$_{0.47}$As/InP are shown
in Fig.\ \ref{fig:InP_slat_fit_avg}.
Again, the error generated by the truncation of the multipole series
vanishes away from the interfaces.  The magnitude of the error is
similar to that in GaAs/AlAs.  For In$_{0.53}$Ga$_{0.47}$As/InP,
however, there is a slight error in the bulk macroscopic electric
fields even before the multipole expansion is truncated.

To test the convergence of the truncated multipole expansions used
here, the power series (\ref{eq:power}) was extended to include terms
up to $l = 6$ in the linear density and up to $l = 4$ in the quadratic
density.  The resulting macroscopic errors (shown in Figs.\
\ref{fig:GaAs_slat_fit_avg} and \ref{fig:InP_slat_fit_avg}) actually
increase slightly, but the change in the Hartree potential remains
negligible.  For the case $\sigma = 6d$ (not shown here), which
includes the most important part of the potential for slowly varying
envelopes, the macroscopic Hartree error of both truncated expansions
is visually almost indistinguishable from that of the complete
quadratic response.

As a direct practical test of the convergence of the multipole
expansion, the extended limits (including also terms of the same order
in the ionic pseudopotential and exchange-correlation potential) were
used to recalculate the superlattice subband structures shown in Sec.\
V of the following paper. \cite{Fore07b} No difference was observable
in any case.  Therefore, the truncated multipole expansions defined in
Eq.\ (\ref{eq:lmax}) are very well converged for slowly varying
envelopes.

\subsection{Bulk energy eigenvalues at $\Gamma$}

\label{subsec:bulkenergy}

The error analysis to this point has focused on the Hartree potential,
with the multipole truncation analysis limited to the macroscopic ($G
= 0$) case.  However, the Hamiltonian for slowly varying envelope
functions is also sensitive to the $G \ne 0$ terms, the
quadratic-response error in the exchange-correlation potential, and
the multipole truncation error in this potential and the local and
nonlocal ionic pseudopotentials.  (There is no quadratic error in the
latter because the ionic pseudopotential is purely linear.)

These sources of error are not analyzed separately here, but to
leading order their net effect is to generate an error in the bulk
energy bands of Figs.\ \ref{fig:GaAs_bulk_bs_AB} and
\ref{fig:InP_bulk_bs_AB}.  This error can be calculated easily at the
$\Gamma$ point by adding linear and quadratic bulk perturbations to
the $\Gamma$ Hamiltonian of the reference crystal, diagonalizing the
resulting matrices, and comparing the approximate $\Gamma$ energies
with the exact values for the given materials.  For example, the
quadratic error in the $\Gamma_{15\mathrm{v}}$ energy is $-0.4$ meV
for GaAs and $+0.5$ meV for AlAs.  Combining these values with the
corresponding bulk macroscopic errors of $+1.9$ meV and $-1.8$ meV for
the Hartree potential in Fig.\ \ref{fig:GaAs_slat_fit_avg}(b), there
is a net error of $+1.5$ meV in the bulk valence-band edge of GaAs and
a net error of 2.7 meV (or 0.6\%) in the GaAs/AlAs valence-band
offset.

Likewise, the quadratic error in the $\Gamma_{15\mathrm{v}}$ energy is
$+3.15$ meV for In$_{0.53}$Ga$_{0.47}$As and $-4.45$ meV for InP\@.
The nominal positions of the interfaces in Fig.\
\ref{fig:InP_slat_fit_avg} are $z = -0.25$ and $z = 23.75$.  The
errors in the macroscopic Hartree potential at the positions halfway
between the interfaces are $+1.35$ meV for In$_{0.53}$Ga$_{0.47}$As
and $-1.32$ meV for InP\@.  Therefore, there is a net quadratic error
of $+4.5$ meV in the bulk valence-band edge of
In$_{0.53}$Ga$_{0.47}$As and a net error of 10.3 meV (or 4.4\%) in the
In$_{0.53}$Ga$_{0.47}$As/InP valence-band offset.  Note that the
quadratic error in each component of the In$_{0.53}$Ga$_{0.47}$As/InP
valence-band offset is only about 1\%, but the total quadratic error
is larger because the Hartree shift and bulk energy shift have
opposite signs, while the corresponding errors have the same signs.

\section{Conclusions}

\label{sec:conclusions}

This paper has investigated numerically a quadratic-response
approximation \cite{Fore05a,Fore05b} for the self-consistent
one-electron potential within a model system based on superlattice LDA
calculations with norm-conserving pseudopotentials.  The electron
density and potential energy of the superlattice were approximated by
retaining only the linear and quadratic response to the
heterostructure perturbation.  This approximation worked well, with a
net absolute error for the $\Gamma_{15}$ valence states of about 2 meV
in GaAs/AlAs and 5 meV in In$_{0.53}$Ga$_{0.47}$As/InP\@.  As shown in
the following paper, \cite{Fore07b} the principal effect of this error
for slowly varying envelope functions is simply a constant shift of
the superlattice energy eigenvalues.

The density and short-range potentials were then approximated further
using truncated multipole expansions (i.e., power series in $k$),
retaining terms of order $k^2$ in the linear potential and $k^0$ in
the quadratic potential.  This had no effect on the macroscopic
density and potential in bulk, but it generated some additional error
(due primarily to the truncation of the linear density response) in a
narrow region near the interfaces.  This error was confirmed to be
negligible for slowly varying envelope functions.

Although the quadratic response is about an order of magnitude smaller
than the linear response, it must be included if the self-consistent
heterostructure potential is to be predicted to an accuracy of better
than a few tens of meV.  The quadratic response is also qualitatively
important because of its different symmetry.  Dipole terms in the
quadratic response were found to produce interface asymmetry and
macroscopic electric fields in the no-common-atom
In$_{0.53}$Ga$_{0.47}$As/InP system.  As shown in the following paper,
\cite{Fore07b} these terms, which have $C_{2v}$ symmetry, produce a
significant fraction of the splitting of the quasidegenerate ground
state in such systems.  Therefore, to provide a realistic description
of such interface-induced symmetry-breaking effects, existing
empirical pseudopotentials would need to be modified to include these
contributions.

\section{Limitations and possible extensions of the theory}

\label{sec:limitations}

\subsection{Quasiparticle effects}

The LDA model system used in the present study is, of course, not
completely realistic because it fails to provide accurate predictions
of the conduction-band states.  This model could be improved by
including quasiparticle self-energies \cite{Aulbur00,Onida02} and by
replacing norm-conserving pseudopotentials with projector augmented
waves \cite{Blochl94,Holz97,Kresse99} (which would provide a better
description of core electrons while remaining within a
pseudopotential-like formalism needed for the applicability of the
present perturbation scheme).  Despite the well-known inaccuracies of
LDA conduction bands, \cite{Mart04} LDA wave functions do have a very
high overlap with $GW$ quasiparticle wave functions. \cite{HybLou86}
This similarity of the wave functions was used by Wang and Zunger
\cite{WaZu95} to construct empirical pseudopotentials with accurate
energy gaps and effective masses using only a small (first-order)
perturbation of the LDA Hamiltonian.  This strongly suggests that,
even though a numerical implementation of quasiparticle self-energies
would be substantially more complicated than the present LDA model,
the basic qualitative conclusion of the present study---namely, that
the quadratic response provides an accurate approximation of the
one-electron heterostructure potential---would remain valid in this
theory also.

\subsection{Atomic relaxation}

Another useful extension of the present study would be to allow the
superlattice ions to relax to their equilibrium positions.  For
example, it is well known that, although In$_{0.53}$Ga$_{0.47}$As and
InP have the same bulk lattice constant, the bond lengths for the InAs
or In$_{0.53}$Ga$_{0.47}$P bonds that form at a heterojunction are
significantly different.  \cite{Hyb90b,PBBR90,DanZun92} The resulting
displacement of the ionic planes from their present ``clamped''
reference-crystal positions would generate additional dipole,
quadrupole, and higher-order moments that have not been included here.
The lowest-order contribution is a dipole term arising from the
displacement of a single plane of ions, which generates a potential
shift \cite{MartKunc81,PBBR90} $\Delta V = 4 \pi Z^* u / A
\epsilon_{\infty}$, where $Z^*$ is the transverse effective ionic
charge, $u$ is the displacement, $A$ is the area per ion in the plane,
and $\epsilon_{\infty}$ is the static electronic dielectric constant.
The qualitative effects of such strain-induced interface dipoles are
the same as those of the purely ``chemical'' dipoles in the
clamped-ion system studied here, but accurate predictions of the
properties of physical heterostructures could only be achieved by
including both contributions.

Atomic relaxation is readily calculated using any of several
structural optimization algorithms already included in the
\textsc{abinit} software package used here.
\cite{Gonze02,Gonze05,ABINIT} The main difficulty in handling strained
superlattices is the same as that encountered in strained bulk
crystals---namely, that the electronic boundary conditions for the
strained system are not the same as those for the unstrained system,
so that the strain cannot be treated by the ordinary methods of
perturbation theory. \cite{BirPik74,Bar01} This difficulty can be
surmounted by an extension \cite{Burt89,WaZu99} of the usual bulk
coordinate-transformation technique \cite{BirPik74} in which the
coordinates of one (or possibly more than one) atom per unit cell in
the strained system are mapped onto the coordinates of the same atom
in the reference crystal.  The coordinate transformation is
straightforward in principle but carries with it an additional heavy
layer of mathematical formalism.  Therefore, in the interests of
simplicity, this complication was excluded by fiat in the present
work, although it may be taken up in future studies.

\subsection{Quantum wires or dots}

The final extension discussed here is the application of the present
methods to quantum wires or quantum dots.  This paper was limited to
superlattices primarily for practical reasons, since, as mentioned in
Sec.\ \ref{subsec:choices}, these calculations are intended to be used
in the following paper \cite{Fore07b} to compare envelope-function
predictions of valence subband structure directly with numerical LDA
results.  Such a direct comparison is possible for superlattices
containing on the order of 100 atoms, which are large enough for
slowly varying envelopes to exist and yet small enough for LDA
plane-wave calculations to be feasible.  However, this clearly would
not be practicable (at present) for a large quantum dot containing
$\sim 10^6$ atoms.

Nevertheless, even if direct comparisons are currently out of reach
for large quantum dots and wires, the quadratic-response approximation
can still be used to construct envelope-function Hamiltonians for such
systems.  Inspection of Fig.\ \ref{fig:InP_dia_den_resp_k} shows that
the response to monatomic and diatomic perturbations is essentially
independent of the size of the supercell as long as the supercell is
large enough for interactions between perturbations in adjacent cells
to be negligible.  Therefore, one can construct the Hamiltonian from
calculations on relatively small supercells containing one or two
atomic perturbations (as opposed to the perturbations comprising one
or two {\em planes} of atoms in the present study).  Indeed, such
calculations have already been performed for monatomic perturbations
in 16-atom fcc supercells in Refs.\ \onlinecite{BaReBaPe89} and
\onlinecite{PBBR90}, although somewhat larger supercells would
probably be necessary for an accurate treatment of diatomic
perturbations.

It should be noted, however, that the one-dimensional multipole
expansions developed in Sec.\ \ref{sec:multipole} for superlattices
are not directly applicable in quantum wires or dots.  For such cases,
one would need to use the more general three-dimensional multipole
expansions developed in Refs.\ \onlinecite{Fore05a} and
\onlinecite{Fore05b}.  These would lead to additional terms in the
envelope-function Hamiltonian, as discussed in Ref.\
\onlinecite{Fore05b} and in footnote 30 of the following
paper. \cite{Fore07b}

\begin{acknowledgments}
This work was supported by Hong Kong RGC Grant No.\ 600905.
\end{acknowledgments}

\appendix*

\section{}

If we consider for simplicity a function of only two $\theta$
variables, the density and potential are written as quadratic
functions
\begin{equation}
  f(\theta_1,\theta_2) = c + c_{1} \theta_1 + c_{2} \theta_2 + c_{11}
  \theta_1^2 + c_{22} \theta_2^2 + 2 c_{12} \theta_1 \theta_2 .
\end{equation}
Here the coefficients $c$, $c_i$, and $c_{ij}$ were calculated by the
direct supercell method. \cite{ResBarBal89} First, the self-consistent
density and potential were found for a supercell consisting of the
reference crystal with a monatomic perturbation $\theta_i = \pm
\Delta_i$ (e.g., in Al$_{0.5}$Ga$_{0.5}$As, one Al$_{0.5}$Ga$_{0.5}$
atom was replaced by either Al$_{0.55}$Ga$_{0.45}$ or
Al$_{0.45}$Ga$_{0.55}$).  This provides the values of the coefficients
\begin{subequations}
\begin{align}
  c & = f_{00} , \\
  c_{1} & = \frac{f_{10} - f_{\bar{1}0}}{2 \Delta_1} , \\
  c_{11} & = \frac{f_{10} + f_{\bar{1}0} - 2 f_{00}}{2 \Delta_1^2} ,
\end{align}
\end{subequations}
where $f_{\bar{1}0} = f(-\Delta_1, 0)$, etc.  Next, diatomic
perturbations were used to find the value of $c_{12}$.  Two possible
formulas were considered:
\begin{subequations}
\begin{align}
  c_{12} & = \frac{f_{11} - f_{1\bar{1}} - f_{\bar{1}1} +
  f_{\bar{1}\bar{1}}}{8 \Delta_1 \Delta_2} , \label{eq:a12_s} \\
  c_{12} & = \frac{f_{11} - f_{10} - f_{01} + f_{00}}{2 \Delta_1
  \Delta_2} . \label{eq:a12_a}
\end{align}
\end{subequations}
The symmetric formula (\ref{eq:a12_s}) is presumably more accurate,
but very little difference was found in comparison to the asymmetric
formula (\ref{eq:a12_a}).  Since the latter requires only one-fourth
the computation time and storage of the former, Eq.\ (\ref{eq:a12_a})
was used for all calculations reported here.

% \bibliography{main}

\begin{thebibliography}{79}
\bibitem[{not({\natexlab{a}})}]{note:MagZun02}
\bibinfo{note}{Although it was not originally expressed in this way, the
  pseudopotential interpolation scheme defined in Eq.\ (A4) of Magri and Zunger
  \cite{MagZun02} is equivalent to the inclusion of diatomic building blocks.}
\bibitem[{not({\natexlab{b}})}]{note:pbcfield}
\bibinfo{note}{If the supercell has a nonvanishing dipole moment, a crystal
  consisting of a finite number of supercells in vacuum does not satisfy
  periodic boundary conditions because of the macroscopic electric field
  (depolarization field) generated by macroscopic polarization charges on the
  surface of the crystal. This depolarization field can be cancelled by
  embedding the crystal in a perfect metal, thus establishing a physical
  context for the use of periodic boundary conditions (see Refs.\
  \onlinecite{BaBaRe88} and \onlinecite{Fraser96}). The macroscopic electric
  field obtained by averaging over the volume of a supercell is therefore
  identically zero under periodic boundary conditions. However, the macroscopic
  electric field obtained by averaging over the volume of a primitive
  zinc-blende unit cell is not everywhere zero. This is apparent in Fig.\
  \ref{fig:InP_dia_den_resp_r}(b), where the macroscopic part of the electric
  field in the regions away from the dipole at the origin is just the field due
  to the induced surface charge on the metal (plus the linear response to this
  field).}
\bibitem[{not({\natexlab{c}})}]{note:macroave}
\bibinfo{note}{At this value of $k$, the Gaussian function in Eq.\
  (\ref{eq:macroave_k}) has a value of $\exp(-\pi/4) \simeq 0.456$.}
\end{thebibliography}

\end{document}